\documentclass[twocolumn]{aastex61}
\usepackage{amsmath}
\usepackage{threeparttable}
\usepackage{float}
\usepackage{graphicx}

\newcommand{\package}[1]{\tt{#1}}
\newcommand{\angstrom}{\textup{\AA}}

\newcommand{\fesc}{$f_{\mathrm{esc}}$}
\newcommand{\fescs}{$f_{\mathrm{esc}}\ $}
\newcommand{\sig}{$\Sigma_{\mathrm{SFR}}$}
\newcommand{\sigs}{$\Sigma_{\mathrm{SFR}}\ $}
\newcommand{\muv}{$M_{\mathrm{UV}}$}
\newcommand{\muvs}{$M_{\mathrm{UV}}\ $}
\newcommand{\ndot}{$\dot{n}_{\mathrm{ion}}$}
\newcommand{\ndots}{$\dot{n}_{\mathrm{ion}}\ $}
\newcommand{\auvs}{$\alpha_{\mathrm{UV}}\ $}
\newcommand{\auv}{$\alpha_{\mathrm{UV}}$}

\shorttitle{Reionization by the Oligarchs}
\shortauthors{Naidu et al.}

\revised{Draft version \today}

\begin{document}

\title{Rapid Reionization by the Oligarchs: The Case for Massive, UV-Bright, Star-Forming Galaxies with High Escape Fractions}

\correspondingauthor{Rohan P. Naidu}
\email{rohan.naidu@cfa.harvard.edu}
\author[0000-0003-3997-5705]{Rohan P. Naidu}
\author[0000-0002-8224-4505]{Sandro Tacchella}
\author[0000-0002-3407-1785]{Charlotte A. Mason}\thanks{Hubble Fellow}
\author[0000-0002-0974-5266]{Sownak Bose}
\affiliation{Center for Astrophysics $|$ Harvard \& Smithsonian, 60 Garden Street, Cambridge, MA 02138, USA}
\author[0000-0001-5851-6649]{Pascal A. Oesch}\affiliation{Department of Astronomy, Universit\'e de Gen\`eve, Chemin des Maillettes 51, 1290 Versoix, Switzerland}\affiliation{Cosmic Dawn Center (DAWN)}
\author[0000-0002-1590-8551]{Charlie Conroy}
\affiliation{Center for Astrophysics $|$ Harvard \& Smithsonian, 60 Garden Street, Cambridge, MA 02138, USA}

\begin{abstract} 
The protagonists of the last great phase transition of the universe -- cosmic reionization -- remain elusive. Faint star-forming galaxies are leading candidates because they are found to be numerous and may have significant ionizing photon escape fractions (\fesc). Here we update this picture via an empirical model that successfully predicts latest observations (e.g., the rapid drop in star-formation density ($\rho_{\mathrm{SFR}}$) at $z>8$). We generate an ionizing spectrum for each galaxy in our model and constrain \fescs by leveraging latest measurements of the reionization timeline (e.g., Ly$\alpha$ damping of quasars and galaxies at $z>7$). Assuming a constant \fescs across all sources at $z>6$, we find \muv$<-13.5$ galaxies need \fesc=$0.21^{+0.06}_{-0.04}$ to complete reionization. The inferred IGM neutral fraction is [0.9, 0.5, 0.1] at $z=[8.2, 6.8, 6.2]\pm0.2$, i.e., the bulk of reionization transpires rapidly in $300$ Myrs, driven by the $z>8$ $\rho_{\mathrm{SFR}}$ and favored by high neutral fractions ($\sim$60$-$90$\%$) measured at $z\sim7-8$. Inspired by the emergent sample of Lyman Continuum (LyC) leakers spanning $z\sim0-6.6$ that overwhelmingly displays higher-than-average star-formation surface density (\sig), we propose a physically motivated model relating \fescs to \sigs and find \fesc$\propto \Sigma^{0.4\pm0.1}_{\mathrm{SFR}}$. Since \sigs falls by $\sim2.5$ dex between $z=8$ and $z=0$, our model explains the humble upper limits on \fescs at lower redshifts and its required evolution to \fesc$\sim0.2$ at $z>6$. Within this model, strikingly, $<$5$\%$ of galaxies with \muv$<-18$ and $\log(M_{\star}/M_{\odot})>8$ (the `oligarchs') account for $\gtrsim$80$\%$ of the reionization budget -- a stark departure from the canonical `democratic' reionization led by copious faint sources. In fact, faint sources (\muv$>$$-$16) must be relegated to a limited role in order to ensure high neutral fractions at $z=7-8$. Shallow faint-end slopes of the UV luminosity function (\auv$>-2$) and/or \fescs distributions skewed toward massive galaxies produce the required late and rapid reionization. We predict LyC leakers like COLA1 ($z=6.6$, \fesc$\sim30\%$, \muv$=-21.5$) become increasingly common towards $z\sim6$ and that the drivers of reionization do not lie hidden across the faint-end of the luminosity function, but are already known to us.
\end{abstract}

\keywords{galaxies: high-redshift ---  galaxies: evolution  --- dark ages, reionization, first stars}

\section{Introduction}
\label{sec:introduction}

The Epoch of Reionization (EoR) marks the last great phase transition of the universe, during which vast islands of neutral Hydrogen were ionized by the first sources of light \citep{Loeb&Barkana01}. The protagonists, topology, and timeline of the EoR are intertwined with our understanding of the early universe and its newly born stellar populations \citep[for a recent review, see][]{Dayal18}. Due to the rapidly fading quasar emissivity at $z>3$ \citep[e.g.,][]{Kulkarni19}, star-forming galaxies are favored to dominate reionization \citep[e.g.,][]{Bouwens15c}. Bright star-forming galaxies have not shown much promise of being effective ionizing sources. Until very recently, these galaxies were measured to have negligible ionizing photon escape fractions \citep[e.g.][]{Steidel18}. This, combined with their observed rarity has meant a reservoir of ultra-faint sources far below current detection limits (modulo highly lensed fields) is widely invoked to drive reionization \citep[e.g.,][]{Livermore17}.

Modeling reionization by star-forming galaxies is typically cast as a tale of three quantities: $\rho_{\mathrm{SFR}}$, $\xi_{\mathrm{ion}}$, and \fescs \citep[e.g.,][]{Madau99,Robertson15,Bouwens15c}. The cosmic star-formation rate density, $\rho_{\mathrm{SFR}}$, provides a measure of star-formation in the early universe. It has now been tracked out to $z\sim10$, with latest studies showing an accelerating drop beyond $z>8$ \citep{Oesch18, Ishigaki18}, consistent with the predictions of simple models that link star-formation rates (SFR) to dark matter accretion \citep[e.g.,][]{Tacchella13,Tacchella18,Mason15,Mashian16}. The ionizing photon production efficiency, $\xi_{\mathrm{ion}}$, is a conversion factor for how many Hydrogen ionizing photons emerge from each episode of star-formation. Tight constraints on $\xi_{\mathrm{ion}}$ can be placed using H$\alpha$ measurements and some assumptions about \fesc, and now exist from direct spectroscopy \citep{Nakajima16,Matthee17,Shivaei18, Tang19} and IRAC-excess inferences of H$\alpha$ out to $z\sim5$ \citep{Bouwens16b, Lam19}. 

The escape fraction, \fesc, is the fraction of ionizing photons generated in a galaxy that evade photoelectric absorption and dust in the Interstellar Medium (ISM) to escape into the neutral Intergalactic Medium (IGM) and ionize it. While $\rho_{\mathrm{SFR}}$ and $\xi_{\mathrm{ion}}$ will be measured ever more precisely with, e.g., the \textit{James Webb Space Telescope (JWST)}, ionizing radiation and thus \fescs will \textit{never} be directly measured in the EoR due to the opacity of the intervening neutral IGM \citep[e.g.,][]{Fan01,Inoue14,McGreer15}. To make things more challenging, it is also extremely difficult to get a handle on \fescs through simulations since it depends sensitively on resolving the multi-phase ISM, and the treatment of small-scale processes associated with galaxy formation, which, at present, are modelled in an approximate way \citep[e.g.,][]{Wise14,Ma15,Trebitsch17}.

However, there is a path forward: since \fescs is by far the single largest uncertainty in modeling reionization, we can employ $\rho_{\mathrm{SFR}}$ and $\xi_{\mathrm{ion}}$ from state-of-the-art measurements and constrain \fescs against latest measurements of the timeline of reionization. The fraction of dark pixels in the Ly$\alpha$ and Ly$\beta$ forests provides a model-independent limit on the end of reionization \citep{McGreer15}. The electron scattering optical depth ($\tau$) reported by \citet[][]{Planck18}, much lower and more precise than previous measures of $\tau$ \citep[e.g.,][]{Hinshaw13}, is an integrated probe of the density of ionizing photons, as CMB photons scatter off of electrons knocked out of Hydrogen atoms. The first quasars and large Ly$\alpha$ surveys at $z\gtrsim7$ allow detailed inferences of the neutral fraction as a function of redshift from the magnitude of Ly$\alpha$ damping \citep[e.g.,][]{Banados18, Mason18}. 

Complementing these data on the global history of reionization are clues about \fescs on a galaxy by galaxy level. For the first time we have a robust sample of individual star-forming galaxies securely detected in Lyman Continuum (LyC) spanning $z\sim0.3-4$ (``LyC leakers") \citep[e.g.,][]{Naidu17,Vanzella18,Rivera-Thorsen19}. The campaigns targeting Green Peas at $z\sim0.3$ with $HST$/COS have proven remarkably efficient \citep[boasting a $100\%$ success rate for LyC detection;][]{Izotov16b,Izotov18b}. Concurrently, a sample at high-$z$ is emerging from deep rest-frame UV spectroscopy \citep[e.g.,][]{Steidel18} and imaging with \textit{HST}/UVIS \citep[e.g.,][]{Fletcher19}. Taken together, these leakers provide hints about the galaxy properties that favor LyC leakage during the EoR. For instance, the overwhelming majority of LyC leakers are compact \citep[e.g.,][]{Izotov18a} and show multi-peaked Ly$\alpha$ \citep[e.g.,][]{Verhamme17}. These insights can be incorporated into models of \fescs that improve upon previous analyses that assumed a single number across the entire galaxy population.

Meanwhile, the bulk of observational constraints on the average LyC \fescs have relied on stacking shallow non-detections for individual galaxies to place stringent upper-limits of \fesc$<10\%$ out to $z\sim4$ \citep[e.g.,][]{Siana10,Rutkowski17, Grazian17, Japelj17, Naidu18}. Taken at face value, these studies effectively rule out an average \fesc$>10\%$ for \muv$\lesssim-20$ sources and put the focus on fainter galaxies, for which no LyC constraints exist yet, as the drivers of reionization. However, if we consider the anisotropy, stochasticity, and evolution with $z$ of \fescs that recent simulations have brought to light \citep[e.g.,][]{Paardekooper15,Trebitsch17,Rosdahl18} along with a higher CGM+IGM opacity, the limits from these studies are far less stringent and can be relaxed by factors of $2-5\times$ \citep[e.g.,][]{Fletcher19, Steidel18}. And indeed, latest studies that emphasize deep spectra and photometry for individual sources find \fesc$\sim10\%$ in stacks of normal $\log(M_{\star}/M_{\odot})\sim8.5-10$ galaxies at $z\sim2.5-4$ \citep{Oesch19, Steidel18, Marchi18}. The emerging observational picture is that average \fescs of $\sim$10$\%$ are possible in relatively bright galaxies (\muv$<-20$).

In parallel, early hydrodynamical simulations produced a mixture of results: with \fescs correlating with halo mass \citep[e.g.,][]{Gnedin08b, Wise&Cen09}, anti-correlating with halo mass \citep[e.g.,][]{Yajima11,Paardekooper15, Kimm17}, and with typical time-averaged values far smaller than \fesc$\sim20\%$ \citep[e.g.,][]{Ma15}. However, contemporary hydrodynamical simulations through a combination of feedback, binaries, turbulence, and careful modeling of the multi-phase ISM are able to produce average \fescs  in the $10-20\%$ range in $\gtrsim10^{8}~M_{\odot}$ galaxies \citep[e.g.,][]{Ma16,Rosdahl18}. \citet{Ma15} and \citet{Ma16} are particularly illustrative of this shift, where these authors first wrote about the difficulty of producing \fescs$>5\%$ due to high absorption in the birth clouds of massive stars but then subsequently found binary models of stellar evolution that destroyed these clouds while retaining highly ionizing sources until late times could achieve \fesc$\sim20\%$.

The driving impulse of this work is to unite the developments outlined above \textit{self-consistently} under the same umbrella to see what story they tell about \fescs and thus reionization. Our umbrella of choice is the empirical galaxy formation model by \citet{Tacchella18} that incorporates recent developments (e.g., cutting-edge stellar population synthesis models) and accurately predicts latest observations (e.g., the sharp drop in $\rho_{\mathrm{SFR}}$) (\S\ref{sec:model_desc}). Leaving \fescs as a free parameter in the equations of reionization (\S\ref{sec:eqns_of_reion}), we fit for it against recently derived constraints on reionization that we describe in \S\ref{sec:data}. Two models of \fescs -- one constant across all galaxies during reionization, another dependent on star-formation surface density -- are justified, set up, and fit in \S\ref{sec:model1} and \S\ref{sec:model2}. The implications of the resulting reionization histories -- their rapid pace, the concentration of the reionization budget among ``oligarch" galaxies, the path forward for observational studies -- are discussed in \S\ref{sec:discussion}. We address open questions and caveats in \S\ref{sec:caveats}. A summary of our findings and an outlook to the future is presented in \S\ref{sec:summary}.

We use \fescs to denote both the singular and plural ``escape fraction" and ``escape fractions". The volume-averaged IGM neutral fraction and ionized fraction are denoted by $\bar{x}_{\mathrm{HI}}$ and $\bar{Q}_{\mathrm{HII}}=1-\bar{x}_{\mathrm{HI}}$ respectively. For cosmological parameters, we adopt the following from \citet{Planck18}: $ h=0.6772$, $X_{\mathrm{p}}=0.75328$, $\Omega_{\mathrm{b}}=0.02241/h^{2}$, $\rho_{\mathrm{c}}=1.8787\times10^{-29}h^2$. All magnitudes are in the AB system \citep{Oke83}.

\section{Methods}
\subsection{An Empirical Model for Galaxy Evolution at $z\geq4$}
\label{sec:model_desc}
The foundation of this work is the empirical model introduced in \citet{Tacchella18}. Here we briefly summarize it, and then describe in detail the quantities relevant to reionization. 

\subsubsection{Model Description}

The \citet{Tacchella18} model is built on top of a $\mathrm{10^{6}~Mpc^{3}}$, high-resolution, N-body, dark-matter simulation, {\sc color} \citep{Sawala16, Hellwing16}. It makes the assumption that the star-formation rate (SFR) of a halo depends on the growth rate of a halo and a star-formation efficiency that is independent of redshift \citep[see also][]{Tacchella13, Mason15}. The halo merger trees self-consistently give rise to star-formation histories for each galaxy from which spectral energy distributions (SEDs) are computed using the Flexible Stellar Population Synthesis code \citep[\texttt{FSPS},][]{fsps1, fsps2, python-FSPS} and the \texttt{MESA} Isochrones and Stellar Tracks which incorporate the effects of rotation \citep[\texttt{MIST},][]{MIST1, MIST2, Choi17}. 

\begin{figure}
\centering
\includegraphics[width=\linewidth]{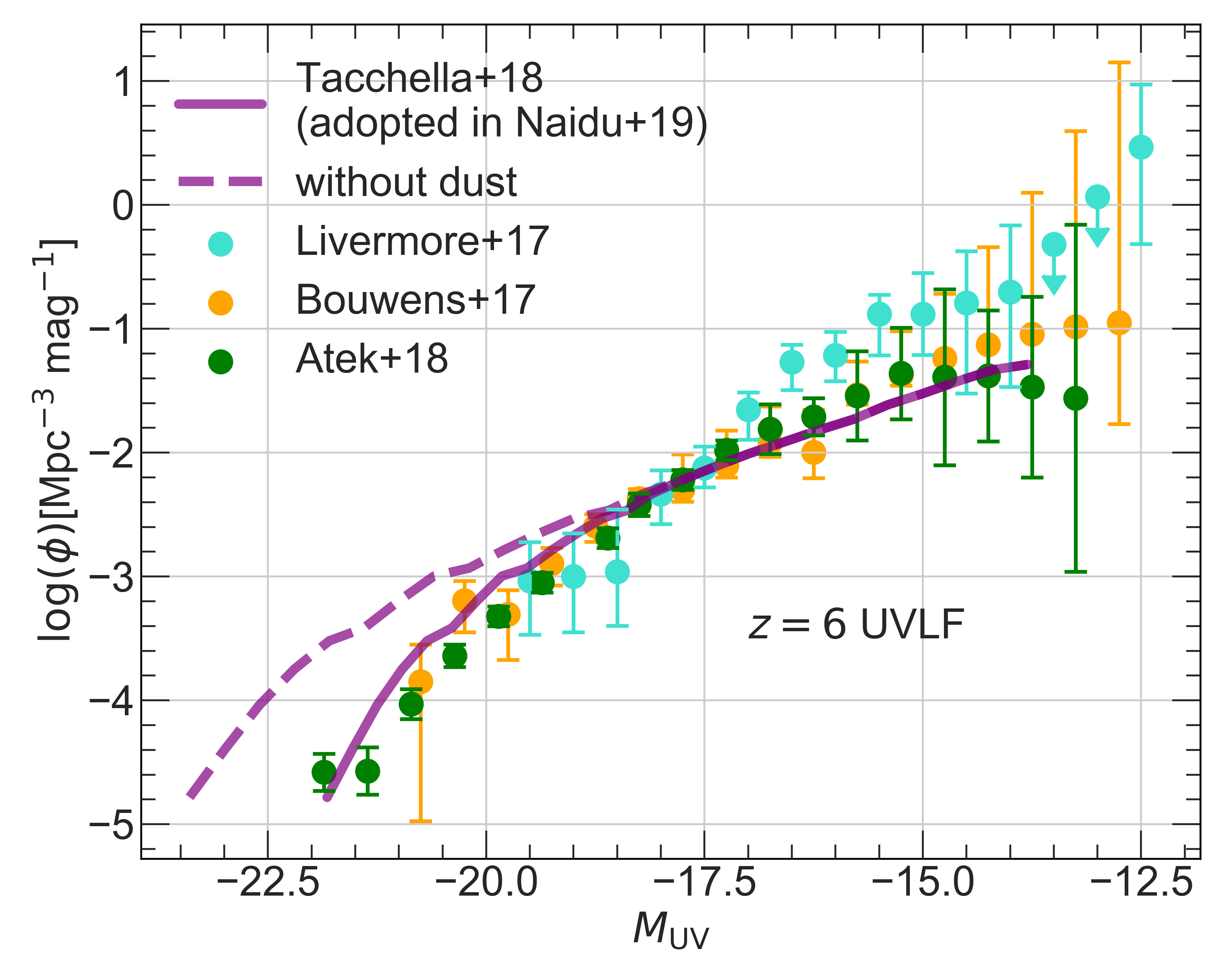}
\caption{The $z=6$ UV Luminosity Function (UVLF). We plot recent determinations using the Hubble Frontier Fields in orange \citep{Bouwens17}, turquoise \citep{Livermore17}, and green \citep{Atek18}. Our predicted UVLF (purple) agrees well with the \citet[][]{Atek18} and \citet[][]{Bouwens17} measurements within errors, and is shallower than the \citet[][]{Livermore17} UVLF. While the formal faint-end slope fit to our UVLF is somewhat shallower than what has been reported, the actual LF is in decent agreement with observations. For this comparison we correct for completeness to account for our box-size ($10^{6}$ Mpc$^{3}$, \citealt[][]{Sheth01}), and also for dust using the UV continuum prescription of \citet[][]{Bouwens14}. Both these corrections only effect the bright-end ($<$$-$20). The \citet{Bouwens17} and \citet{Livermore17} points are adjusted by 0.15 dex to account for differences in their mean redshift following \citet{Atek18}.}
\label{fig:uvlf}
\end{figure}

The star-formation efficiency of the model is tuned to match the $z=4$ Ultraviolet Luminosity Function (UVLF) and then predicts UVLFs out to $z\sim10$ consistent with the observed data (see Figure \ref{fig:uvlf} in this work and Figure 3 in \citealt{Tacchella18}). The faint-end slope of the UVLF (\auv) in our model steepens with redshift and the same trend is seen in observations \citep[e.g.,][]{Finkelstein15,Bouwens15aLF}. For our best-fit Schechter function at $z=$[4, 6, 8, 10, 12] we find $\alpha_{\mathrm{UV}}=$[$-1.63\pm0.02$, $-1.72\pm0.03$, $-1.69\pm0.04$, $-1.84\pm0.06$, $-1.99\pm0.24$], consistent with recent empirical and semi-analytical models that find $\alpha_{\mathrm{UV}}=-1.5$ to $-2.0$ at $z=5-10$ \citep[e.g.,][]{Yung19,Endsley19,Behroozi19}. Our best-fit faint-end slope values are somewhat shallower than what is reported in the recent literature from observations \citep[][]{Bouwens17,Livermore17,Atek18}. However, we note the Schechter function parameters are highly degenerate and that our slopes are fit over a different range of \muvs (we truncate at \muvs$<-13.5$) compared to observations. Figure \ref{fig:uvlf} shows the actual LF from our model is in decent agreement with the latest observations at $z\sim6$, down to the faintest limits that can currently be probed in the Hubble Frontier Fields. We provide UVLFs for our model out to $z=12$ and $\chi^{2}$ comparisons against literature data in Appendix \ref{appendix:uvlf}. Note that we compute the ionizing budget via SEDs of individual sources, and not by integrating under the UVLF. We discuss $\alpha_{\mathrm{UV}}$ in detail in \S\ref{sec:democratic}.

The resolution of the dark-matter simulation limits our model to $\mathrm{SFR\ga 0.02\ M_{\odot}/yr}$, corresponding to $\mathrm{M_{UV}\lesssim -13.5}$, roughly where estimates of the faint end of the UVLF begin to diverge due to magnification uncertainties of the lensing models \citep[e.g.,][]{Bouwens17}. In all calculations we integrate down to this limit. Our fiducial model adopts a \citet{Salpeter55} IMF, a constant metallicity of $\mathrm{ Z_{\star}/Z_{\odot}=0.02}$, and the UV continuum dust prescription of \citet{Bouwens14} (their Table 3). Swapping the \texttt{MIST} models for those that explicitly include binaries \citep[\texttt{BPASS},][]{bpass}, or a \citet{Chabrier03} IMF, or a model with evolving metallicity make no appreciable difference to our results. Dust, the faint-end slope of the UVLF, and the effect of changing the \muvs cutoff are discussed in \S\ref{sec:caveats}.

\subsubsection{The Ionizing Photon Production Efficiency ($\xi_{\mathrm{ion}}$)}
\label{sec:xi_ion}

\begin{figure}
\centering
\includegraphics[width=0.95\linewidth]{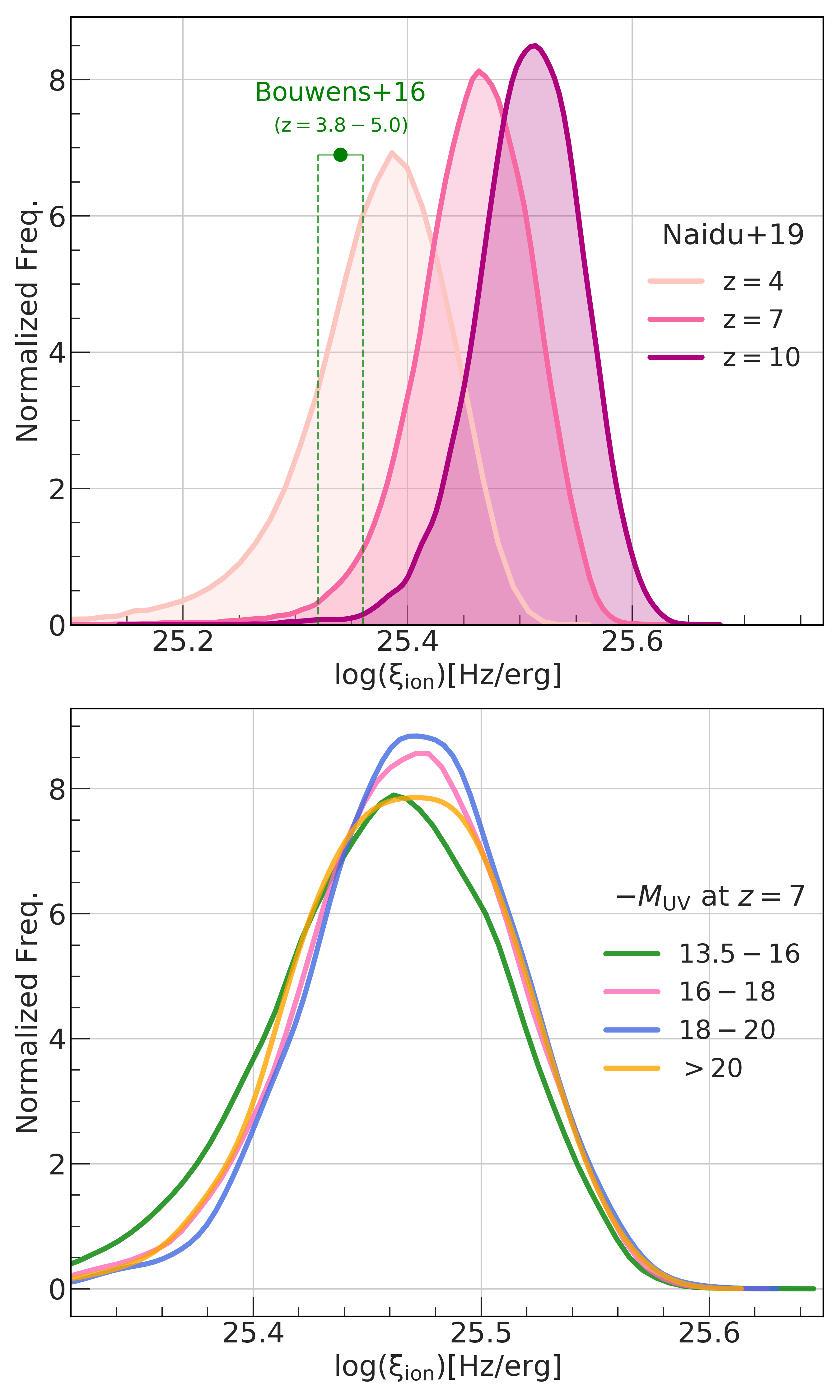}
\caption{The ionizing photon production efficiency ($\xi_{\mathrm{ion}}$) predicted by our model using \texttt{FSPS}+\texttt{MIST}. $\xi_{\mathrm{ion}}$ represents the number of ionizing photons produced in a galaxy before being absorbed by the ISM or attenuated by dust. \textbf{Top:} At $z\sim4-5$, the highest redshift at which $\xi_{\mathrm{ion}}$ has been measured statistically, our model agrees within error-bars with the \citet{Bouwens16b} estimate (shown in green). We predict evolution in $\xi_{\mathrm{ion}}$ as the stellar populations grow older, with a $40\%$ smaller median at $z=4$ than at $z=10$. \textbf{Bottom:} At fixed redshift we predict $\xi_{\mathrm{ion}}$ does not vary with the brightness of galaxies.}
\label{fig:emissivity_zevol}
\end{figure}

$\xi_{\mathrm{ion}}$ provides a measure of the gross LyC photons produced at a given time in a source. It is typically cast in terms of the rate of ionizing photons ($\mathrm{N(H^{0})}$) per unit UV luminosity, usually measured at $1500\angstrom$ ($\mathrm{L_{1500}} $):

\begin{equation}
    \xi_{\mathrm{ion}} = \frac{N(H^0)}{L_{1500}} [\mathrm{s^{-1}/erg\  s^{-1}Hz^{-1}}].
\end{equation}

We compute $\xi_{\mathrm{ion}}$ for each galaxy in the empirical model directly from its SED by integrating the flux produced below the Lyman limit to obtain $N(H^{0})$ and then  normalizing by the SED-flux at $\mathrm{1500\angstrom}$. The \texttt{MIST} isochrones that our SEDs are synthesized from include the effects of rotation that boost the ionizing flux production of massive stars akin to, but not exactly like, the effect of binaries \citep{Choi17}. The harder UV spectra produced by these rotating models (or binaries) are the only kind that self-consistently explain the strong nebular fluxes, high ionization lines, and extreme line widths that are now known to be ubiquitous at $z\sim2.5-3.5$ \citep[e.g.,][]{Steidel16,Holden16,Reddy18b} and also commonly seen at $z>6$ \citep[e.g.,][]{Roberts-Borsani16,Stark15,Mainali17}.

In Figure \ref{fig:emissivity_zevol} we show the $\log_{10}(\xi_{\mathrm{ion}})$ distribution for the galaxies in our model. We predict the median $\xi_{\mathrm{ion}}$ between $z=4-10$ rises by $\sim40\%$ ($\sim 0.15$ dex) as galaxies get younger at higher redshift and their ionizing spectra become harder. We compare our predictions with \citet{Bouwens16b}, who report a mean $\log_{10}(\xi_{\mathrm{ion}}) = 25.34^{+0.02}_{-0.02}\ (25.54^{0.12}_{-0.12}) $ for a sample spanning $z=3.8-5.0 \ (5.0-5.5)$ using an SMC attenuation curve, which \citet{Reddy18} show to be the appropriate curve for sub-solar metallicity populations expected at $z>4$. At $z=4$ ($z=5$) we have a median $\mathrm{log_{10}(\xi_{\mathrm{ion}}) = 25.37^{+0.06}_{-0.06} \ (25.40^{+0.05}_{-0.06}}$) that agrees with their measurements within error-bars. At fixed redshift we find $\xi_{\mathrm{ion}}$ does not vary with mass or \muv. \citet{Lam19} observe a similar invariance with brightness at $z\sim4-5$ for \muv$<-17.5$ galaxies. These trends hold even when using \texttt{BPASS} templates and with an evolving metallicity. While not directly dealing with the redshifts that are the focus of this work, we note that the $\xi_{\mathrm{ion}}$ values reported for $z\sim2-3$ galaxies are in broad agreement with our model assuming a linear extrapolation to lower redshifts \citep{ Nakajima16, Matthee17, Shivaei18}.

\begin{figure}
\centering
\includegraphics[width=1.05\linewidth]{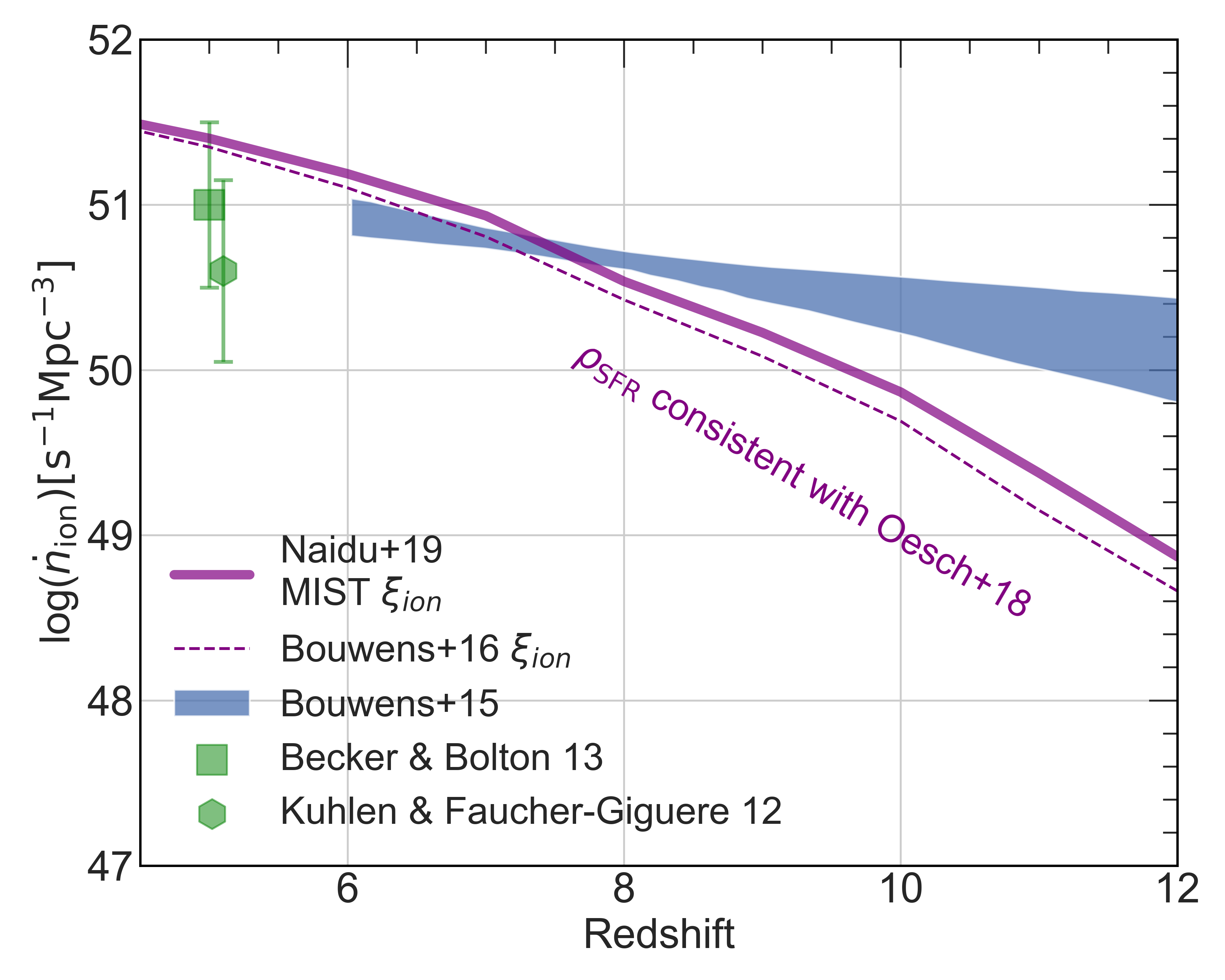}
\caption{$\dot{n}_{\mathrm{ion}}$, the co-moving emissivity of ionizing photons (Eq. \ref{eq:nion}), as a function of redshift. Shown in blue is the model from \citet{Bouwens15c}, representative of fits based on the \citet{Planck2015} $\tau$. Our model, here set to \fesc$=0.2$, \muv$<-13.5$, and with \auv$\gtrsim-2$ predicts a sharp drop at $z>8$ consistent with the latest \textit{HST} $\rho_{\mathrm{SFR}}$ \citep{Oesch18, Ishigaki18}. At $z=10$, our model produces $\sim10\times$ fewer ionizing photons with the difference disappearing at $z\lesssim7$. This dearth of LyC in the early universe, as we show later (Figure \ref{fig:fourpanel}), compresses the timeline of reionization. The \texttt{MIST} models with rotation produce higher $\xi_{\mathrm{ion}}$ (solid purple) than \citet[][dashed purple; comparable to the \citealt{Robertson15} $\xi_{\mathrm{ion}}$]{Bouwens16b}, but nowhere close to bridging the gulf between the purple and blue curves. Shown in green are $z\sim4-5$ Ly-forest measurements of $\dot{n}_{ion}$ \citep{Becker13,Kuhlen12}.}
\label{fig:csfrd}
\end{figure}

Typically, reionization studies set $\xi_{\mathrm{ion}}$ to some fixed, redshift-invariant value that lies in the locus our galaxies span in Figure \ref{fig:emissivity_zevol}. This is a reasonable assumption as evidenced by the narrow spread ($\sim0.1$ dex at each redshift across $z=4-10$) and gradual evolution during reionization ($z=6-10$). What this means is that when considering the \textit{total} integrated ionizing output at a particular redshift, there is not much of a distinction between our approach and assuming some reasonable fixed value (note how close the dotted and dashed purple lines are in Figure \ref{fig:csfrd}). However, the advantage of our model is that it captures the diversity in $\xi_{\mathrm{ion}}$ on a galaxy by galaxy basis so that we are able to link \fescs to \textit{individual} galaxy properties and through the product \fesc$\times N(H^{0})$ probe how much each galaxy contributes to reionization (see also \S\ref{sec:model2}, Figure \ref{fig:AOE}, and Table \ref{table:AOE}).

\subsubsection{Cosmic Star-Formation Rate Density ($\mathrm{\rho_{SFR}}$)}

At $z=4-8$ the \citet[][]{Tacchella18} model is in excellent agreement with the consensus $\mathrm{\rho_{SFR}}$ \citep[e.g.,][]{Bouwens15aLF, Finkelstein15, madau&dickinson}. At $z>8$, where various measurements diverge, the model predicts a drop in $\rho_{\rm SFR}$ consistent with the latest \textit{HST} analyses from \citet{Ishigaki18} and \citet{Oesch18}. The sharp drop in $\rho_{\mathrm{SFR}}$ in our model comes as the bulk of halos at $z>8$ begin to fall below the halo-mass corresponding to maximal star-formation efficiency ($\mathrm{M_{h}\sim 10^{11}-10^{12}~ M_{\odot}}$). 

The difference between earlier smooth power-law fits for $\mathrm{\rho_{SFR}}$ at $z>4$, which use steeper $\alpha_{\mathrm{UV}}\lesssim-2$ (e.g., $\mathrm{\rho_{SFR}\propto} (1+z)^{-4.2}$; \citealt{Bouwens15aLF, Finkelstein15, Robertson15}) and our model, which predicts an accelerating decline in $\mathrm{\rho_{SFR}}$, is as large as an order of magnitude at $z=10$ and three orders at $z=14$. This difference is directly reflected in the dearth of LyC photons available for reionization at early times. Earlier works \citep[e.g.,][]{Robertson15} had already shown that reionization likely proceeds without significant contribution from $z>10$ sources. The dearth of sources in our model (and thus LyC) at $z\sim8-10$ combined with other data, as we shall see in \S\ref{sec:rapid}, even further compresses the timeline of reionization and pushes it to later times.

\subsection{Equations of Reionization}
\label{sec:eqns_of_reion}
We closely follow the widespread approach that models reionization as an interplay between ionization and recombination \citep[e.g.,][]{Madau99, Robertson13}. Here we outline the relevant equations.  

We start with the quantity directly inherited from our empirical model: the co-moving production rate of hydrogen-ionizing photons ($\dot{n}_{ion}$), i.e., the gross number of LyC photons escaping into the IGM per unit time per unit volume. $\dot{n}_{ion}$ is usually computed as follows:

\begin{equation}
\label{eq:nion}
    \dot{n}_{\mathrm{ion}}(z) = f_{\mathrm{esc}}\xi_{\mathrm{ion}}\rho_{\mathrm{UV}_{\mathrm{dust\ corr}}}\  [\mathrm{s^{-1}Mpc^{-3}}],
\end{equation}

\noindent
where $\rho_{\mathrm{UV}_{\mathrm{dust\ corr}}}$ is the dust-corrected UV luminosity.

This UV-anchored formulation, where $f_{\mathrm{esc}}$ is the escape fraction of ionizing photons relative to the dust-corrected observed UV, is apt for working with observations, where $\rho_{\mathrm{UV}}$ is the measured quantity around which all else is based. However, in our model which is built on a $\mathrm{10^{6}~Mpc^{3}}$ simulation box, we simply sum the LyC photons, $N(H^{0})$, produced by every galaxy from their SEDs directly, reducing Equation \ref{eq:nion} to:

\begin{equation}
\label{eq:n_dot}
    \dot{n}_{\rm ion}(z) = \sum_{M_{\mathrm{UV}}<-13.5} \frac{f_{\rm esc}N(H^{0})}{10^6}\  [\mathrm{s^{-1}Mpc^{-3}}].
\end{equation}

Here $f_{\rm esc}$ is the escape fraction of ionizing photons relative to the total ionizing photons produced in the galaxy. 

The IGM ionized fraction, $\bar{Q}_{\mathrm{HII}}$, is evolved as per the following differential equation where the first term represents ionization, and the second recombination,

\begin{equation}
\label{eq:Q_dot}
    \dot{Q}_{\mathrm{HII}}= \frac{\dot{n}_{\rm ion}}{\langle n_{\mathrm{H}}\rangle} - \frac{\bar{Q}_{\mathrm{HII}}}{t_{\mathrm{\rm rec}}},
\end{equation}

\noindent
where $\langle n_{\mathrm{H}}\rangle = X_{\mathrm{p}}\Omega_{\mathrm{b}}\rho_{\mathrm{c}}$ is the co-moving density of Hydrogen, which depends on the primordial mass-fraction of Hydrogen ($X_{\mathrm{p}}$), the fractional baryon density $\Omega_{\mathrm{b}}$ and the critical density $\rho_{\mathrm{c}}$. $t_{\mathrm{rec}}$, the recombination time of ionized Hydrogen in the IGM, is given by
\begin{equation}
\label{eq:t_rec}
    1/t_{\mathrm{rec}}=C_{\mathrm{HII}}\alpha_{\mathrm{B}}\left(1+\left(1-X_{\mathrm{P}}\right)/4X_{\mathrm{P}}\right)\langle n_{\mathrm{H}} \rangle \left(1+z\right)^{3},
\end{equation}
where $C_{\mathrm{HII}}=\langle n_{\mathrm{H}}^{2} \rangle/\langle n_{\mathrm{H}} \rangle^{2}$ is the clumping factor that models the inhomogeneity of the IGM which we set to 3, and $\alpha_{\mathrm{B}}=2.6\times10^{-13}\left(T/10^4\mathrm{K}\right)^{0.76}\mathrm{cm^3s^{-1}}$ is the Case B recombination coefficient at electron temperature, $T$ that we set to $10^4\mathrm{K}$ \citep{Shull12, Robertson13,Pawlik15, Robertson15, SF16}.

The Thomson optical depth, $\tau$, is calculated as 

\begin{equation}
\label{eq:tau}
    \tau(z) =  c\langle n_{\mathrm{H}} \rangle \sigma_{T} \int_{0}^{z} f_{\mathrm{e}} \bar{Q}_{\mathrm{HII}}H(z)^{-1}\left(1+z'^{2}\right)dz',
\end{equation}

\noindent
where $\tau$ is the Thomson optical depth, $c$ is the speed of light, $\sigma_{T}$ is the Thomson scattering cross-section, $f_{e}$ is the number of free electrons for every Hydrogen nucleus in the ionized IGM that we set to $\left(1+(1-X_{\mathrm{P}})/4X_{P}\right)$ \citep{Kuhlen12}, and $H(z)$ is the Hubble parameter.

We note there are caveats: for example, Case B recombination may not be an appropriate description towards the end of reionization when local absorption in dense clumps becomes important \citep{Furlanetto05}, or $C_{\mathrm{HII}}$ is bound to evolve as the universe grows more ionized \citep{Pawlik15} though its effect on reionization inference is limited \citep{Mason19nonpar,Bouwens15c}. While we could test the effect of assuming different values for every individual parameter (on top of varying the IMF, metallicity, underlying SED models, and the truncation $M_{\mathrm{UV}}$), our guiding philosophy is to hew to the canonical assumptions from the literature so all our divergent conclusions are clearly attributable to the new data we constrain against and the models we introduce in this work.

The only free parameter in these equations is the escape fraction, \fesc. Our strategy to constrain \fesc, and thus constrain reionization histories is to solve Equations \ref{eq:n_dot}-\ref{eq:tau} assuming a model for \fescs and then fit the model parameters against the data described in the following section.

\subsection{Observational Constraints on Reionization}
\label{sec:data}
Here we enumerate state-of-the-art measurements on the timeline of reionization that we use to constrain our \fescs models. We briefly describe each measurement and specify how it is included in our inference.

\begin{enumerate}
    \item {Thomson Optical Depth:} $\tau=0.0540\pm0.0074$ from \citet{Planck18}, which is a more precise, downward revision of $\tau=0.066\pm0.012$ \citep{Planck16} and a far cry from the WMAP $\tau=0.088\pm0.014$ \citep{Hinshaw13}. $\tau$ bears the imprint of free electrons on photons from the last-scattering surface of the CMB and provides a global, integrated, model-independent constraint. The lower value of $\tau$ from \citet{Planck18} allows for the sharp drop in $\rho_{\mathrm{SFR}}$ at $z>8$ (Figure \ref{fig:csfrd}) which was disfavored by earlier measurements \citep[e.g.,][]{Robertson13}.
    
    \item {Ly$\alpha$, Ly$\beta$ dark fraction:} $\bar{x}_{\mathrm{HI}} \leq0.06\pm0.05$ at $z=5.9$, as per the model-independent ``dark fraction" in Ly$\alpha$ and Ly$\beta$ forests of quasar spectra \citep{Mesinger10,McGreer15}. The completely dark pixels in the forests are either due to neutral \ion{H}{1} in the IGM and/or astrophysical interlopers and hence provide an assumption-free upper limit to the global neutral fraction. This constraint allows us to impose the ``end of reionization" in a more self-consistent fashion than abrupt truncation at some redshift (often fixed to $z=6$ \citep[e.g.,][]{Planck16, Planck18}). We adopt it as uniform for $\bar{x}_{\mathrm{HI}}<0.06$ and as a half-Gaussian peaked at $0.06$ with $\sigma=0.05$ elsewhere \citep[][\S 3.1]{Greig17}. 
    
    \item {$z\sim7-8$ Ly$\alpha$ Equivalent Width (EW) Distributions:} $\bar{x}_{\mathrm{HI}}=0.59^{+0.11}_{-0.15}$ at $z\sim7$, $\bar{x}_{\mathrm{HI}}=0.88^{+0.05}_{-0.10}$ at $z\sim7.5$, and $\bar{x}_{\mathrm{HI}}>0.76$ at $z\sim8$ \citep{Mason18, Hoag19, Mason19}. The Ly$\alpha$ line -- in particular, its EW and its velocity offset from the systemic redshift -- bears the imprint of the neutral IGM that \citet{Mason18} infer using empirical fits for the ISM, and state-of-the-art IGM and Ly$\alpha$ radiative transfer simulations \citep{Mesinger16}. While the evolution in the fraction of Ly$\alpha$ emitters in Lyman-break galaxies \citep[e.g.,][]{Mesinger15} encodes the evolution of $\bar{x}_{\mathrm{HI}}(z)$, \citet{Mason18} show that a more competitive constraint may be derived by utilizing EW distributions. We use their full posterior PDF on $\bar{x}_{\mathrm{HI}}$ (their Figure 11) and adopt scatter in the redshift at which they report $\bar{x}_{\mathrm{HI}}$ as per the selection function for their sample -- centred on $z=6.9$ and  $\sigma_{z}= 0.5$ \citep{Grazian12, Pentericci14}, which is consistent with the scatter for similarly selected \textit{z}-band dropouts \citep[e.g.,][]{Bouwens15aLF}. \citet{Mason19} and \citet{Hoag19} apply the same technique at higher redshifts and we adopt their measurements in similar fashion.
    
    \item {$z>7$ Quasi-stellar Objects (QSOs):} $\bar{x}_{\mathrm{HI}}=0.48^{+0.26}_{-0.26}$ at $z=7.09$ and $\bar{x}_{\mathrm{HI}}=0.60^{+0.20}_{-0.23}$ at $z=7.54$ \citep{Davies18b} from the IGM Ly$\alpha$ damping wing signature \citep{Miralda-Escude98} in the quasars ULAS J1120+0641 \citep{Mortlock11} and ULAS J1342+0928 \citep{Banados18}. This constraint arises from a similar approach as \citet{Mason18} in that, detailed empirical models \citep{Davies18a}, IGM simulations \citep{Mesinger11} and radiative transfer \citep{Davies16b} inform the inference of $\bar{x}_{\mathrm{HI}}$ from quasar spectra. We adopt their conservative PDF for $\bar{x}_{\mathrm{HI}}$ (their Figure 11).
    
    \item {$z=6-7$ Ly$\alpha$ Emitter (LAE) Fraction:} The drop in number-density of LAEs between $z\sim6$ and $z\sim7$ may be interpreted as the universe undergoing drastic evolution in neutrality between these redshifts \citep{Mesinger15} but may also be due to survey incompleteness at faint magnitudes \citep{Oyarzun17}. \citet{Greig17} conservatively quantify this as implying $\bar{x}_{\mathrm{HI}}(z=7)- \bar{x}_{\mathrm{HI}}(z=6) \geq 0.4$ and we adopt their weak half-Gaussian constraint peaked at $\bar{x}_{\mathrm{HI}}=1$ with $\sigma=0.6$. Note that the dark fraction constraint at $z=5.9$ along with the $z\sim7$ Ly$\alpha$ EWs already effectively reproduce this sharp change in neutrality.
    
    \item {$z\sim6.6$ Ly$\alpha$ Emitter Clustering:} The observed LAE clustering function at $z\sim6.6$ \citep{Ouchi10} when interpreted in the context of the clustering in detailed reionization simulations suggests $\bar{x}_{\mathrm{HI}}<0.5$ \citep{Sobacchi14} which we implement as a half-Gaussian peaked at zero with $\sigma=0.5$ \citep{Greig17}.
    
    \end{enumerate}
    
\begin{figure*}
\centering
\includegraphics[width=0.95\linewidth]{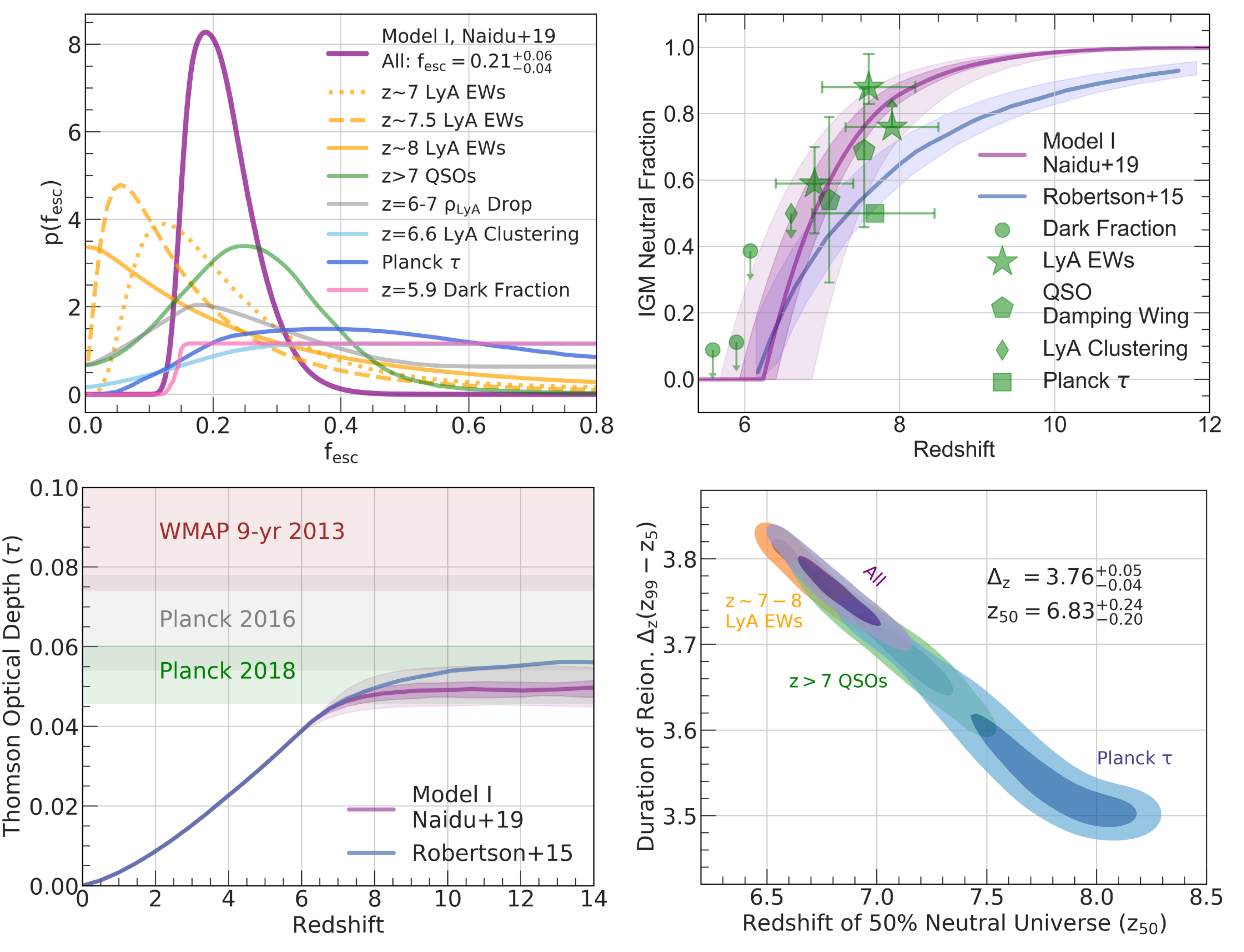}
\caption{Summary of our fits to Model I, in which we assume a constant $f_{\mathrm{esc}}$ for all galaxies at $z>6$. {Top Left:} The allowed $f_{\mathrm{esc}}$ parameter space implied by the reionization constraints described in \S\ref{sec:data}. The model-independent \textit{Planck} $\tau$ (blue) and $z=5.9$ dark fraction (pink) rule out $f_{\mathrm{esc}}\lesssim10\%$, while the $z\sim7$ Ly$\alpha$ profiles (orange) and $z>7$ QSOs (green) are most constraining. The resulting $f_{\mathrm{esc}}=0.21^{+0.06}_{-0.04}$ during reionization requires evolution in \fescs from $\sim10\%$ at $z=3$ and $\sim0\%$ at $z\sim1$ \citep[e.g.,][]{Siana10, Steidel18}. {Top Right:} The evolution of $\bar{x}_{\mathrm{HI}}$, the IGM neutral fraction. The most likely reionization history is tracked in purple (1 and 3$\sigma$ bounds shaded). Literature inferences of the neutral fraction are plotted in green (see \S\ref{sec:data}). In the mean, reionization starts later and proceeds faster than what earlier constraints suggested \citep[e.g.,][shown in blue]{Robertson15} or what the \textit{Planck} $\tau$ alone implies (green square). \textbf{Bottom Left:} The evolution of the Thomson Optical Depth, $\tau$. Our model's drop in ionizing emissivity at $z>8$ (Figure \ref{fig:csfrd}) and thus lower $\tau$ (purple) were previously disfavored by WMAP (brown strip) and earlier \textit{Planck} results (grey strip). However, the latest \textit{Planck} $\tau$ (green strip) allows for it. \textbf{Bottom Right:} The duration of reionization in redshift-space against $z_{50}$, the redshift of the $50\%$ neutral universe. We find tight bounds on both $z_{\mathrm{50}}$ and $z_{\mathrm{99}}-z_{\mathrm{5}}$ combining all our constraints, while $\tau$ by itself is only sensitive to $z_{\mathrm{50}}$ \citep[e.g.,][]{Trac18}.
The blue contours representing $\tau$ come from the $\tau$-$f_{\mathrm{esc}}$ distribution (top left panel), and are not directly inherited from \textit{Planck} -- they derive $z_{50}=7.64\pm0.74$ while we favor even later reionization with $z_{50}=6.83^{+0.24}_{-0.20}$.}
\label{fig:fourpanel}
\end{figure*}

We exclude some measurements: Gamma Ray Burst (GRB) damping spectra, while probing some of the highest redshifts \citep{Chornock13, Totani06, Tanvir09}, preferentially arise out of low-mass halos \citep{Savaglio09}. Given the extreme scatter along the lines of sight to such halos across a patchy IGM, bounds on $\bar{x}_{\mathrm{HI}}$ are fated to be weak \citep{McQuinn08}. Ly$\alpha$-based measurements that do not vary the intrinsic line-width \citep[e.g.,][]{Ouchi10, Inoue18} are likely optimistic, given the vast diversity of ISM conditions in galaxies evident in line-shapes already seen in $z\sim2-3$ samples \citep[e.g.,][]{Trainor15, Trainor16}.

In what follows we set up and justify two models for \fescs that we fit against all the constraints described in this Section.

\begin{figure}
\centering
\includegraphics[width=1\linewidth]{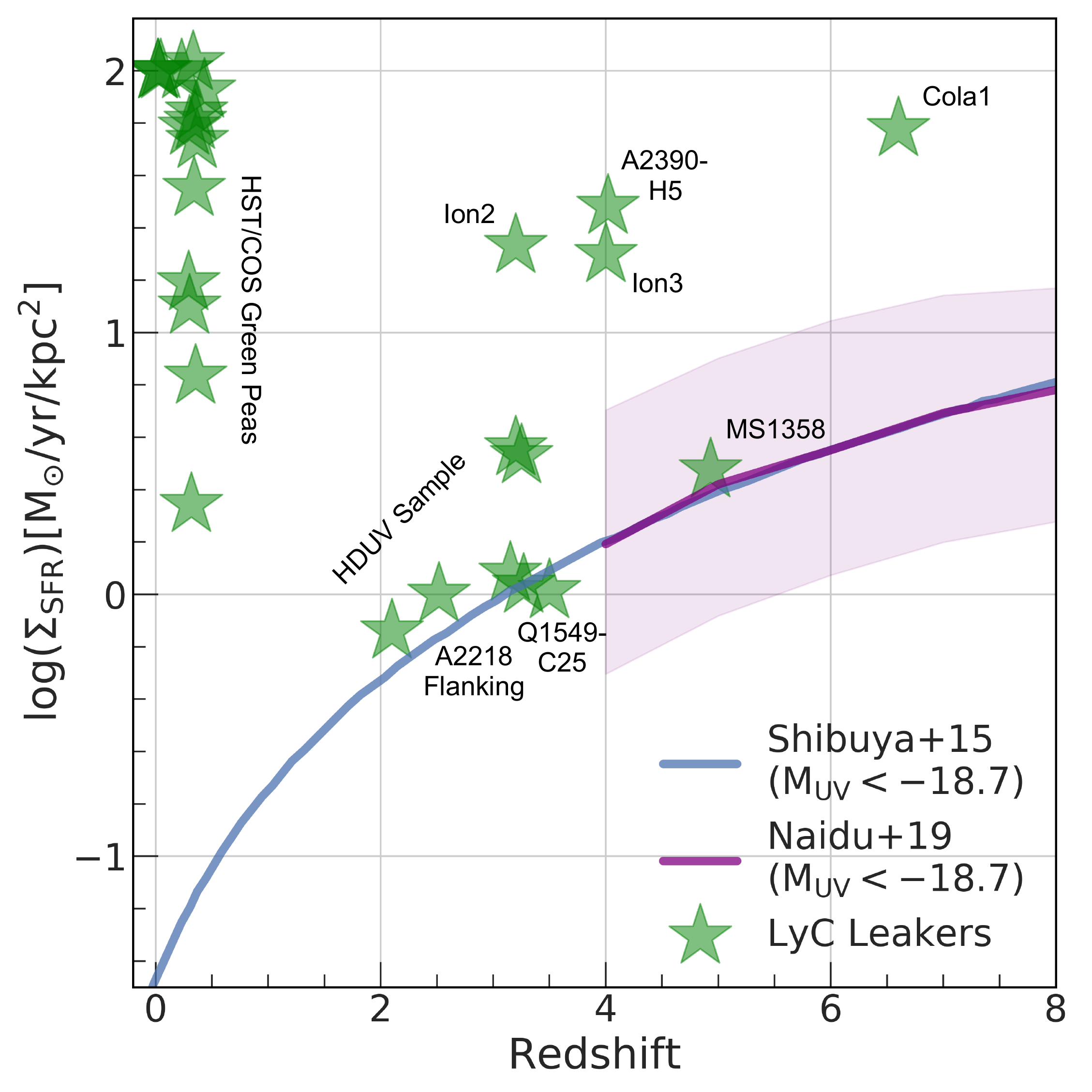}
\caption{The redshift-evolution of star-formation rate surface density ($\Sigma_{\mathrm{SFR}}$).
Our model (purple, $1\sigma$ shaded) remarkably matches the observed $\Sigma_{\mathrm{SFR}}$ shown in blue \citep[][]{Shibuya15} at $z>4$ by setting only a single parameter, $\lambda=R_{\mathrm{vir}}/R_{\mathrm{halo}}=0.031$ which defines the normalization. $\Sigma_{\mathrm{SFR}}$ grows by $\gtrsim2$ dex between the local universe and the Epoch of Reionization. Motivated in part by \textit{almost all} confirmed LyC leakers to date (green stars) showing higher $\Sigma_{\mathrm{SFR}}$ than the average at their redshifts, in our Model II we link $f_{\mathrm{esc}}$ to $\Sigma_{\mathrm{SFR}}$ (see \S\ref{sec:model2}). The green stars represent leakers presented by \citet{Naidu17,deBarros15, Vanzella16, Vanzella18, Shapley16, Bian17, Matthee18, Borthakur14, Izotov16a, Izotov16b, Izotov18a, Izotov18b, Leethochawalit16, Jones13}; Naidu et al. (in prep.) and their vertical error-bars, typically $<$0.2 dex, are omitted and log($\Sigma_{\mathrm{SFR}}$) is capped to 2 for clarity. At $z\sim0$, the aggressive bunching of LyC leakers in the top-left is due to the \citet{Izotov16b} selection that successfully targeted Green-Pea galaxies with \textit{HST}/COS. Between $z\sim2-4$ \textit{HST}/F275W, \textit{HST}/F336W and ground-based UV spectrographs (e.g., \textit{Keck}/LRIS) come into play. Finally at $z\geq4$ when the IGM becomes opaque to LyC, indirect methods must be invoked (e.g., Ly$\alpha$ line-profiles, covering fraction of low-ionization gas).}
\label{fig:sfrsd_zevol}
\end{figure}

\section{Fitting for \lowercase{$f_{\mathrm{esc}}$} Model I: Constant \lowercase{$f_{\mathrm{esc}}$} During Reionization}
\label{sec:model1}
Here we assume the $f_{\mathrm{esc}}$ of \textit{all} galaxies during reionization to be a constant number and denote this as ``Model I". Effectively, we fit for a single normalization factor, \fesc, that sets the scale of the emissivity (solid curve in Figure \ref{fig:csfrd}). This is the common approach adopted in several reionization studies \citep[e.g.][]{Robertson15, Ishigaki18}. Model I ignores the diversity of galaxies and the highly likely dependence of \fescs on various galaxy properties. However, this simple model provides a useful benchmark for the ``average" escape fraction that observational stacking studies compute. Further, intrinsic galaxy properties (e.g., sizes, average star-formation rates) evolve modestly between $z=6-10$ where the bulk of reionization is expected to occur, hence assuming a constant average is justified.

We assume a uniform prior between 0 and 1 on \fescs and depict the resulting posteriors projected in various spaces in Figure \ref{fig:fourpanel}. Combining all constraints we find \fesc=$0.21^{+0.06}_{-0.04}$. Simply requiring reionization to be mostly complete by $z=5.9$ via the dark fraction rules out the \fesc$\lesssim15\%$ parameter space (upper left panel of Figure \ref{fig:fourpanel}). \fesc$\lesssim15\%$ is also disfavored by the \textit{Planck} $\tau$. Note that as the dark fraction and $\tau$ are model-independent constraints not much can be invoked to allow for \fesc$\lesssim15\%$ (we discuss $M_{\mathrm{UV}}$ truncation and the faint-end slope of the UVLF in \S\ref{sec:muv_trunc} and \S\ref{sec:democratic}). The most constraining measurements on \fescs prove to be from the damping wing analysis of quasars and Ly$\alpha$ EW distributions which both require significant neutral fractions at later times ($\bar{x}_{\mathrm{HI}}\sim0.5$ at $z\sim7$). 

In this constant \fescs model we make no claims about the \fescs at $z<6$ -- our result is situated in the reionizing universe. \fesc$=0.21$ is larger than the negligible \fescs measured in deep stacks at $z\sim0-1$ \citep[e.g.,][]{Siana10,Rutkowski16}, where the IGM does not impede observations, and the recently established \fesc$\sim10\%$ at $z\sim2.5-4$ \citep[][]{Oesch19, Marchi17,Steidel18, Fletcher19}. 

To self-consistently bridge these findings of \fesc$\sim0\%$ at $z\sim0$, \fesc$\sim10\%$ at $z\sim2$, and \fesc$\sim20\%$ at $z>6$ we introduce Model II below, which accounts for an evolving \fescs while also considering the diversity in properties of individual galaxies.

\section{Fitting for \lowercase{$f_{\mathrm{esc}}$} Model II: \lowercase{\fesc} as a function of \sig}
\label{sec:model2}
Here we propose a model where \fescs for each galaxy is solely dependent on its star-formation rate surface density $\Sigma_{\rm SFR}$, \fesc = $a\times\Sigma_{\mathrm{SFR}}^{b}$ (where $a$ and $b$ are free parameters which we fit). We justify why this is an apt formulation, specify how it is implemented in our empirical model, and discuss the constraints it yields.

\subsection{Motivation: Why $\Sigma_{\mathrm{SFR}}$?}

Almost all the individual observed LyC leakers to date spanning $z\sim0-6.6$ show $\Sigma_{\mathrm{SFR}}$ higher than the average $\Sigma_{\mathrm{SFR}}$ expected at their redshifts. We demonstrate this in Figure \ref{fig:sfrsd_zevol} where we have compiled all galaxies for which convincing LyC leakage is reported, and that have sizes and UV SFRs available. These include the \textit{HST}/COS sample at $z\lesssim0.3$ \citep{Borthakur14, Heckman11, Izotov16a, Izotov16b, Izotov18a, Izotov18b}, the \textit{HST}/F336W, \textit{HST}/F275W, and ground-based UV spectrograph sources at $z\sim2-4$ \citep[][Naidu et al. in prep.]{deBarros15, Vanzella16, Shapley16, Bian17, Naidu17, Vanzella18}, and sources that show strong indirect hints of LyC escape at $z\sim4-6.6$ (low covering fractions: \citealt{Jones13, Leethochawalit16}, tightly-spaced, double-peaked Ly$\alpha$ resembling the local Green-Pea sample: \citealt{Matthee18}). The SFRs are all calculated from the UV because the average $\Sigma_{\mathrm{SFR}}$ vs. $z$ relation we calibrate our model against is derived from the UV \citep{Shibuya15}. A caveat is that the striking abundance of sources populating the top-left corner of Figure \ref{fig:sfrsd_zevol} were \textit{selected} to be Green-Pea-like \citep[e.g.,][]{Izotov16b} for further follow-up, i.e., with very high \sig, but it is nonetheless remarkable that the selection is so successful given the long history of LyC non-detections. While these individual LyC sources provide useful clues about the properties favoring LyC escape, they may be extreme outliers given their rarity. However, \citet{Marchi18} find that even among normal star-forming galaxies at $z\sim4$, the UV compact sources (which hence have higher \sig) are likelier to be leaking LyC.

Independently, recent state-of-the-art hydrodynamical simulations have put forth the scenario of spatially concentrated star-formation, turbulence, and feedback carving out channels in the ISM through which LyC photons can stream out of the galaxy \citep[e.g.,][]{Ma16,Sharma16,Safarzadeh16, Trebitsch17,Katz18, Rosdahl18,Kakiichi19,Kimm19}. For instance, \citet{Ma16} describe supernovae clearing ionized channels in the ISM around stellar birth-clouds. However, the massive stars exploding as supernovae are precisely the ones producing most of the ionizing photons. Invoking effects like binarity and rotation allow a significant population of UV luminous stars to survive longer and pump LyC through the newly cleared ISM \citep{Choi17}. The ionized channels visible in high-resolution Ly$\alpha$ spectra of LyC leakers \citep{Vanzella19} support this picture.

\subsection{$\Sigma_{\mathrm{SFR}}$ in our Empirical Model}
We use the usual definition of \sigs \citep[e.g.,][]{Shibuya19}:
\begin{equation}
    \Sigma_{\mathrm{SFR}}=\frac{\mathrm{SFR}/2}{\pi R_{\mathrm{gal}}^{2}}.
\end{equation}

The SFR in our model is a function of the halo accretion rate and a redshift-invariant efficiency that converts the halo accretion rate into an SFR. To calculate effective radii ($R_{\mathrm{gal}}$) for the galaxies in our model we assume the angular momenta of the galaxies are a fixed fraction of their DM halo \citep{Mo98}. In particular, we relate $R_{\mathrm{gal}}=\lambda R_{\mathrm{halo}}$ where $\lambda$ is the spin parameter of the halo. We set $\lambda = 0.031$ to reproduce the observed \sig-$z$ relation from \citet{Shibuya15} (see Figure \ref{fig:sfrsd_zevol}). Remarkably, we are able to match the exact evolution of \sigs with redshift via this single parameter that only sets the normalization. This is in line with \citet[][]{Shibuya15}'s finding that the ratio of galaxy size and halo size does not evolve significantly with redshift. We add log-normal scatter while assigning sizes, $\sigma_{\mathrm{log\lambda}}=0.22$ dex consistent with \citet{Kravtsov13, Somerville18, Jiang18}.

\subsection{Fitting for \fesc $\propto$ \sigs}

\begin{figure}
\centering
\includegraphics[width=1\linewidth]{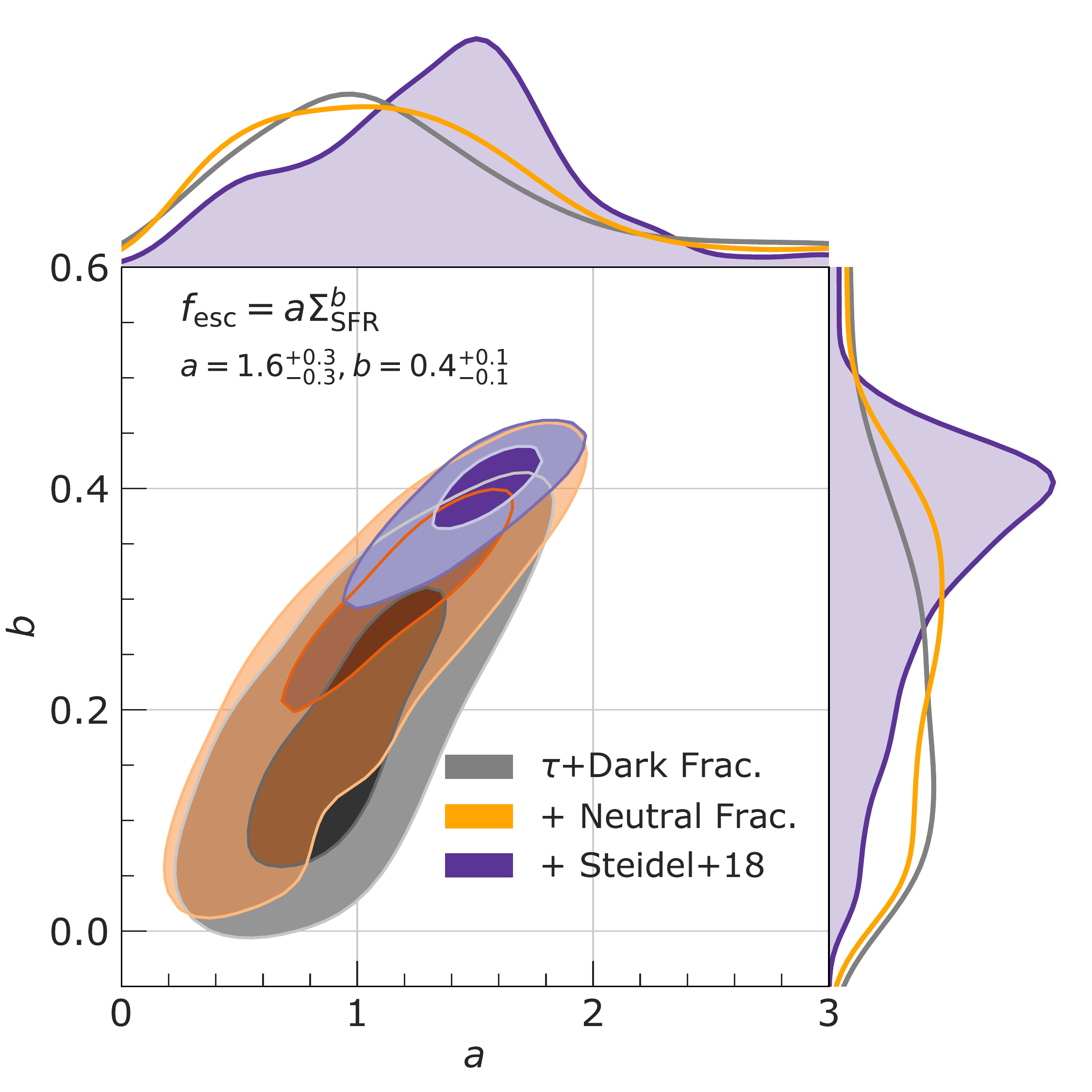}
\caption{Relative constraining power of various data on Model II. While the global constraints on the evolution of the neutral fraction and $\tau$ (gray and orange) produce a degenerate surface, the fit is constrained by the \citet{Steidel18} measurement (purple) of \fescs that we adopt by assuming the \sigs for their sample follows the average relation in Figure \ref{fig:sfrsd_zevol}. We adopt uniform priors of 0 to 5 for $a$ and $-5$ to 5 for $b$, allowing for a \sig-\fescs anti-correlation that is rejected by the evidence. The posteriors are instructive for future studies: complex models of \fescs only constrained by global quantities like $\bar{x}_{\mathrm{HI}}$ and $\tau$ result in degenerate parameters (gray and orange above) and require measurements like \citet[][]{Steidel18} (purple) that directly link \fescs to galaxy properties.}
\label{fig:model2constraint}
\end{figure}

\begin{figure*}
\centering
\includegraphics[width=0.95\linewidth]{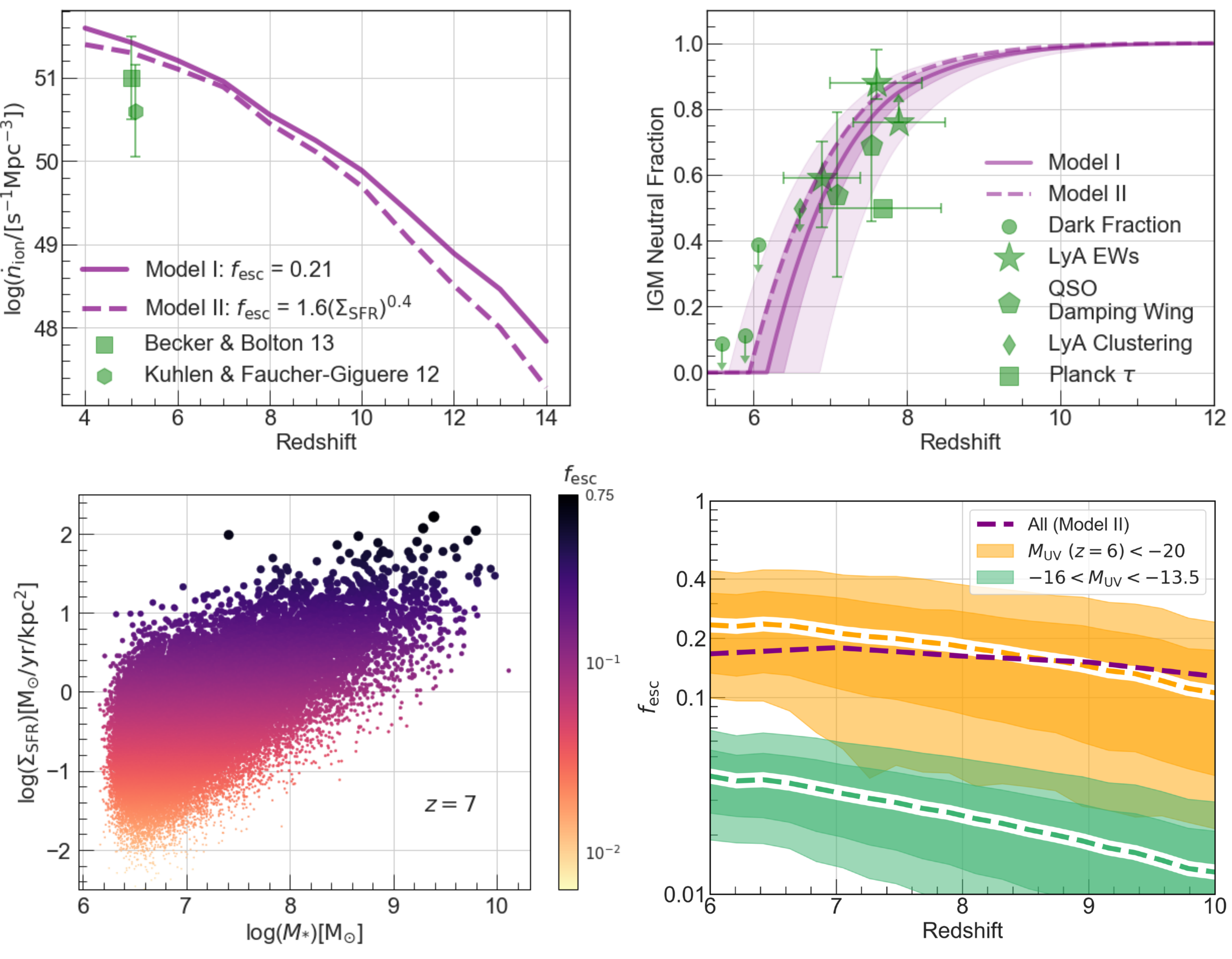}
\caption{Summary of our fits to Model II, where we find \fesc$\propto$\sig$^{0.4}$. The global reionization histories produced by Model II are very similar to those from Model I as seen through the evolution of the ionizing emissivity (\textbf{top left}) and the IGM neutral fraction (\textbf{top right}). However, the distribution of \fescs among galaxies differs significantly. The \textbf{bottom left} panel plots \sigs as a function of stellar mass at $z=7$ with points colored and sized according to their \fescs (larger points denote higher \fesc). Galaxies at $\log(M_{\star}/M_{\odot})\sim8-10$ achieve significantly higher \fescs than lower mass galaxies, though note the large scatter and that many of these galaxies are also able to attain \fescs$>10\%$. In the \textbf{bottom-right} we show the evolution of the mean \fescs (dashed) for the UV brightest (orange) and faintest (green) galaxies with 16th/84th and 5th/95th percentiles shaded. Faint galaxies with very low \sigs are limited to a mean \fesc$<10\%$ (though note the large scatter in the bottom-left panel) while the brightest galaxies are at \fesc$\sim20\%$. The mean \fescs across \textit{all} galaxies (purple) remains a fairly flat $\sim20\%$ akin to Model I and as expected from the similar evolution in \ndots seen in the top-left panel.}
\label{fig:model1vmodel2}
\end{figure*}

We assume a simple power law dependence \fesc$= a\times\left(\Sigma_{\mathrm{SFR}}/\Sigma_{\mathrm{SFR, max}}\right)^{b}$. $\Sigma_{\mathrm{SFR, max}}=1000\  \mathrm{M_{\odot} yr^{-1} kpc^{-2}}$ is close to the value for the maximum \sigs that can be sustained without radiation pressure instabilities \citep{Thompson05SFRSD, Heiderman10, Hopkins10}. The scatter in $\lambda$ produces a maximum \sigs of typically $\sim220\ \mathrm{M_{\odot} yr^{-1} kpc^{-2}}$ in our model. We fit for the coefficients $a$ and $b$ by summing the ionizing photon contributions of each individual galaxy as detailed in \S\ref{sec:eqns_of_reion} and perform Bayesian inference against the reionization constraints from \S\ref{sec:data} using the \texttt{dynesty} nested sampling package \citep{Speagle19}.

\begin{figure*}
\centering
\includegraphics[width=0.95\linewidth]{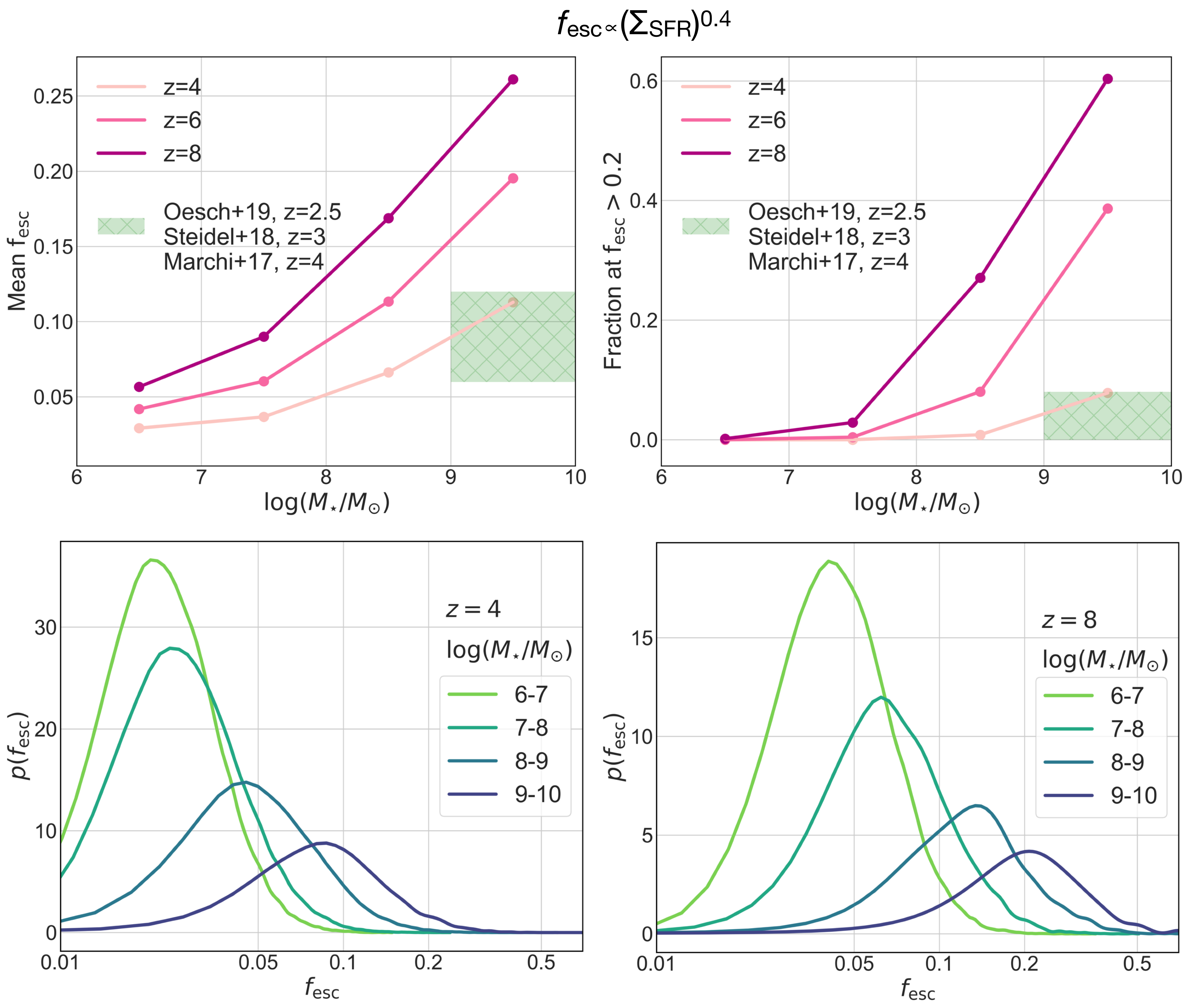}
\caption{The evolution of \fescs as a function of stellar mass and redshift from Model II (\fesc$\propto \Sigma^{0.4}$). \textbf{Top Left:} The mean \fescs at fixed stellar mass grows with redshift as galaxies grow more compact and star-forming though the \fescs of the lowest mass galaxies remains negligible even at $z=8$. The mean \fescs in the highest mass bins at $z\sim8$ reaches $\sim25\%$ and at $z=4$ it is comparable with current constraints at $z=2.5-4$ on ``normal" star-forming galaxies (green hatched region). \textbf{Top Right:} The fraction of galaxies with \fescs above the mean during reionization ($\gtrsim20\%$) shows similar trends. This is consistent with the current observational situation at $z\sim2-4$ (green hatched region) where a small fraction of sources like \textit{Ion2} ($\log(M_{\star}/M_{\odot})\sim9$) show high \fesc, even $>50\%$, while mean stacks (top-left) find humble estimates. We predict the fraction of \textit{Ion2}-like galaxies grows strongly at fixed mass. \textbf{Bottom:} \fescs probability densities at $z=4$ (left) and $z=8$ (right) summarized in the top panels. The key features are the rightward shift of the distributions with increasing $z$ and the high \fescs tails in the right panel corresponding to the ``oligarchs".}
\label{fig:fescfrac}
\end{figure*}

\begin{figure*}
\centering
\includegraphics[width=0.95\linewidth]{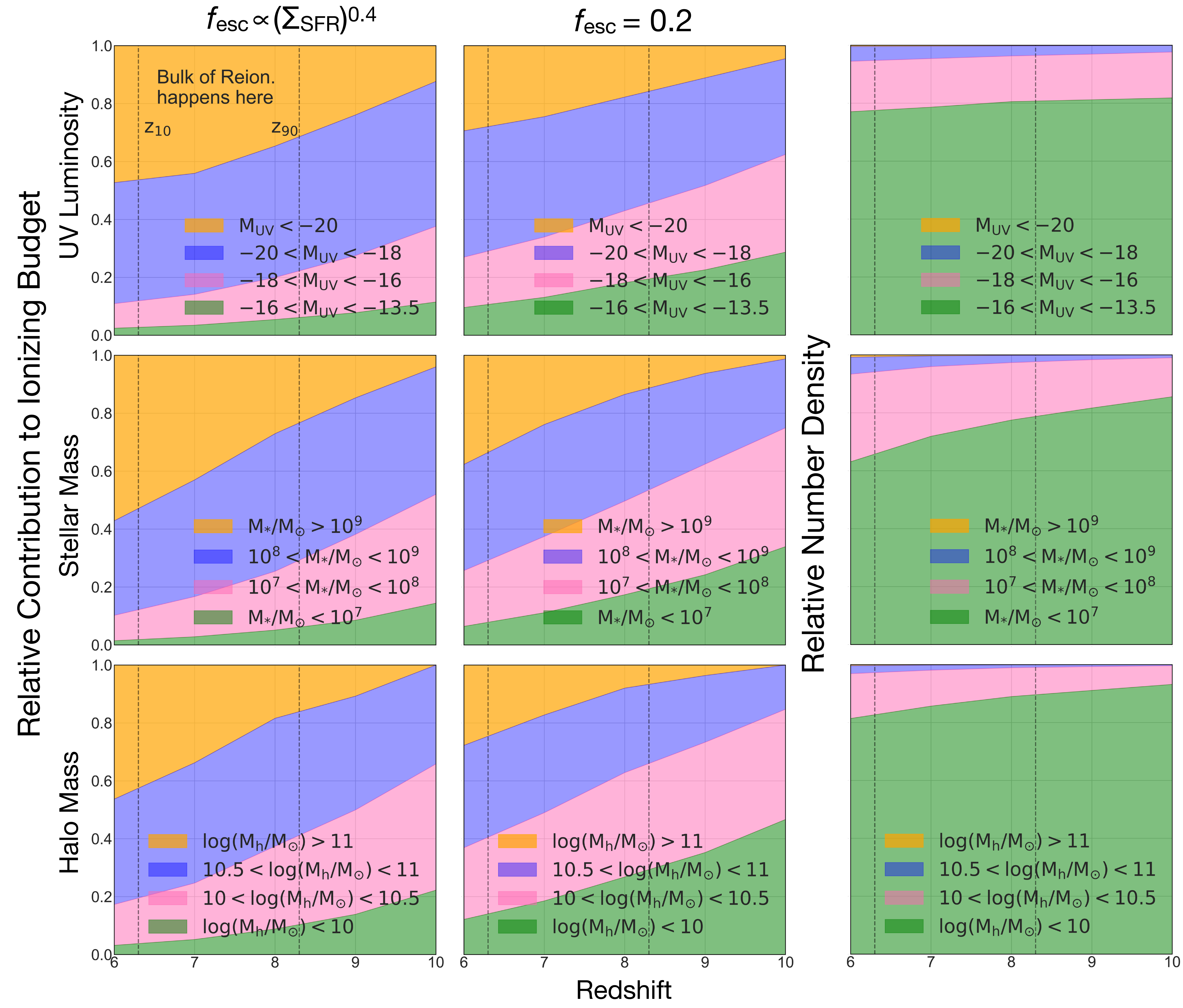}
\caption{Which galaxies reionized the universe? The top, middle, and bottom panels show groupings of galaxies by $\mathrm{M_{UV}}$ (observed), stellar mass, and halo mass respectively. The left and center columns are shaded by the relative contribution of each group to the total ionizing budget ($\dot{n}_{ion}$) as per our two models for \fesc. The right-most column is colored by the relative number-density of each group. The black dotted lines in each panel sandwich the redshift-space when the universe is inferred to go from $90\%$ ($z_{90}$) to $10\%$ neutral ($z_{10}$). In the central panel, the reionization budget is essentially a reflection of our predicted UVLF since $\xi_{\mathrm{ion}}$ does not vary strongly with luminosity and \fescs is constant, while in the left panel the \fesc$\propto$\sigs model further down-weights the contribution of faint galaxies. UV-bright and massive sources represented in orange and blue dominate the reionization budget ($\gtrsim50-80\%$) despite comprising $\lesssim5\%$ of the population. This scenario, ``reionization by oligarchs", stands in stark contrast to the canonical ``democratic reionization" led by copious faint sources.}
\label{fig:AOE}
\end{figure*}

We add one additional constraint: the \fescs of the \citet{Steidel18} sample, \fesc$=0.09\pm0.01$ for a stack of $\sim120$ $M_{\mathrm{UV}}<-19.5$ LBGs that we assume follow the average \sig-$z$ relation at $z\sim3$. While it is impossible to robustly constrain \fescs for individual sources at high-$z$ due to the stochasticity of the intervening IGM, \citet{Steidel18} stack in a narrow redshift bin across multiple lines of sight and correct for the mean IGM at that redshift. Further, the extremely deep spectra in their sample ($\sim10$-hour exposures on a 10m telescope) show weak ISM lines that can be used to fine-tune models to match the covering fraction, correct for attenuation, and produce a robust estimate of \fesc. The key assumption here is that the relationship between \sigs and \fescs at $z\sim3$ holds at higher-$z$ -- we argue that since \fescs largely depends on the covering fraction of neutral gas at $z>3$, and not dust, this is a justifiable assumption (see \S\ref{sec:dust}). We do not include any of the individual LyC leakers depicted in Figure \ref{fig:sfrsd_zevol} in our fits because estimates of \fescs for any individual source are highly uncertain due to the transmission along a single IGM line of sight being unmeasurable.

We find $a=1.6^{+0.3}_{-0.3}$ and $b=0.4^{+0.1}_{-0.1}$ by deploying a uniform prior over 0 to 5 for $a$ and $-5$ to 5 for $b$. We assume a uniform prior such that $0\leq$\fesc$\leq1$, so our best-fit relation effectively is:

\begin{equation}
    f_{\mathrm{esc}} = min\left(1, 1.6^{+0.3}_{-0.3}  \times \left(\frac{\Sigma_{\mathrm{SFR}}}{\Sigma_{\mathrm{SFR, max}}}\right)^{0.4^{+0.1}_{-0.1}} \right).
\end{equation}

The tight posteriors are driven by the \citet{Steidel18} constraint that directly links \fescs to \sig, while the constraints from \S\ref{sec:data} are useful in deciding the positive sign of the dependence (Figure \ref{fig:model2constraint}). We emphasize that in fitting for this power-law we allow for negative powers (i.e., an  \fesc-\sigs anti-correlation) that are rejected by the evidence since they fail to conclude reionization by $z\sim6$.

Model I fits for a very similar evolution of the ionizing photon budget, $\dot{n}_{\mathrm{ion}}(z)$, compared to our more physically motivated Model II (top panels of Figure \ref{fig:model1vmodel2}). Which is to say, the evolution of $\dot{n}_{\mathrm{ion}}(z)$ and average \fescs of $\sim20\%$ in both the models is similar during reionization. However, the way the similar $\dot{n}_{\mathrm{ion}}(z)$ is distributed among galaxies differs radically between the two models in that instead of a constant \fesc$=0.2$ across all galaxies, a minority of galaxies that are more massive and UV bright tend to have high \sigs and thus high escape fractions (bottom panels of Figure \ref{fig:model1vmodel2}). The proportion of these high \sigs galaxies as well as the mean \sigs grows with redshift driven primarily by their increasing compactness and naturally explains the evolution of \fescs from $\sim0\%$ at $z\sim0$ to $\sim10\%$ at $z\sim2.5-4$ to $\sim20\%$ at $z>6$ (Figure \ref{fig:fescfrac}).

\begin{table*}[th]
\centering
\caption{Which galaxies reionized the universe? Mean properties weighted by contribution to reionization ($\dot{n}_{\mathrm{ion}}$) for Model II (Model I).}
\label{table:AOE}
\begin{tabular}{lccccc}
\hline
\multicolumn{1}{l}{Parameter} & \multicolumn{1}{c}{$z=6$} &
\multicolumn{1}{c}{$z=7$} &
\multicolumn{1}{c}{$z=8$} &
\multicolumn{1}{c}{$z=9$} &
\multicolumn{1}{c}{$z=10$}\\
\multicolumn{1}{l}{} & 
\multicolumn{5}{c}{$\mathrm {f_{esc} = 1.62\times \Sigma_{SFR}^{0.42}}$ ($\mathrm{f_{esc}}$=0.21)}\\
\hline
\noalign{\smallskip}
$\mathrm{f_{esc}}$ & 0.24 (0.21) & 0.26 (0.21) & 0.24  (0.21) & 0.21 (0.21) & 0.19 (0.21)\\
$\mathrm{M_{UV}\tablenotemark{$a$}}$ & -19.7 (-18.9) & -19.5 (-18.6) & -19.2 (-18.2) & -18.7 (-17.8) & -18.3 (-17.3)\\
$\mathrm{log(M_{\star}/M_{\odot})}$ & 9.5 (9.2) & 9.2 (8.9) & 8.9 (8.6) & 8.6 (8.4) & 8.3 (8.0)\\
$\mathrm{log(M_{halo}/M_{\odot})}$ & 11.1 (10.9) & 10.9 (10.7) & 10.7 (10.6) & 10.5 (10.4) & 10.4 (10.2)\\
$\mathrm{SFR/M_{\odot}yr^{-1}}$ & 15.3 (9.3) & 12.1 (6.5) & 7.7 (4.0) & 3.5 (2.1) & 2.0 (1.0)\\
$\mathrm{\Sigma_{SFR}/M_{\odot}yr^{-1}kpc^{-2}}$ & 17.7 (8.2) & 23.2 (9.7) & 17.6 (8.1) & 11.6 (5.5) & 9.3 (4.3)\\
$\mathrm{R_{gal}/kpc}$ & 0.59 (0.62) & 0.46 (0.48) & 0.38 (0.39) & 0.33 (0.34) & 0.27 (0.27)\\
\noalign{\smallskip}
\end{tabular}
\begin{tablenotes}
\item[](a) This is the observed $\mathrm{M_{UV}}$, i.e., we have applied attenuation to the intrinsic $\mathrm{M_{UV}}$ as described in \S\ref{sec:model_desc}.
\end{tablenotes}
\end{table*}

\section{Discussion}
\label{sec:discussion}

\subsection{Rapid Reionization at $z=6-8$}
\label{sec:rapid}
Both our models produce virtually identical reionization histories (top right panel of Figure \ref{fig:model1vmodel2}). The mid-point of reionization ($z_{50}$) is $6.83^{+0.24}_{-0.20}$ while the duration of reionization ($z_{90}-z_{10}$) is tightly constrained to be $\Delta_{z}=3.76^{+0.05}_{-0.04}$ (bottom right panel, Figure \ref{fig:fourpanel}). The universe goes from $90\%$ neutral at $z=8.22^{+0.25}_{-0.22}$ to $10\%$ neutral at $z=6.25^{+0.26}_{-0.22}$, in $\sim300$ Myrs (see Table \ref{table:rapid}). This pace is faster than estimated by earlier studies \citep[e.g.,][]{Robertson15, Planck18} and is driven by the sharp drop in the ionizing emissivity at early times (Figure \ref{fig:emissivity_zevol}). The high neutral fraction measurements at late times ($\gtrsim50\%$ at $z\sim7$) from Ly$\alpha$ damping combined with the dark fraction requirement for the end of reionization by $z\sim5.9$ favor this rapid pace. A corollary of this timeline is that efforts to understand the sources of reionization do not have to probe the highest redshifts since more than half of the process occurs between $z=6-7$.

\begin{table}[t]
\centering
\caption{A Rapidly Reionizing Universe}
\label{table:rapid}
\begin{tabular}{lcc}
\hline
\multicolumn{1}{l}{Redshift} & \multicolumn{1}{c}{ $\bar{x}_{\mathrm{HI}}$ (Model I)} & \multicolumn{1}{c}{ $\bar{x}_{\mathrm{HI}}$ (Model II)}\\
\hline
\noalign{\smallskip}
$\mathrm{6}$ & $0.00_{-0.00}^{+0.05}$ & $0.01_{-0.01}^{+0.08}$\\
$\mathrm{7}$ & $0.58^{+0.08}_{-0.12}$ & $0.64_{-0.04}^{+0.03}$\\
$\mathrm{8}$ & $0.87^{+0.03}_{-0.04}$ & $0.89_{-0.01}^{+0.01}$\\
$\mathrm{9}$ & $0.95^{+0.01}_{-0.01}$ & $0.97_{-0.01}^{+0.01}$\\
$\mathrm{10}$ & $0.99^{+0.01}_{-0.01}$ & $0.99_{-0.01}^{+0.01}$\\
\noalign{\smallskip}
\end{tabular}
\end{table}

\subsection{The `Oligarchs' that Reionized the Universe}
\label{sec:oligarchs}
In both our models, especially in Model II, we find the reionization budget (\ndot) is concentrated in a ultra-minority of galaxies with high \sigs at \muv$<-18$, $\log(M_{\star}/M_{\odot})>8$ (see Figure \ref{fig:AOE} and Table \ref{table:AOE}). In Model II less than $5\%$ of galaxies constitute $\gtrsim80\%$ of the reionization budget. Adopting a popular income inequality measure from Macroeconomics, the Gini coefficient \citep{Gini12}, we find the \ndots distributions for Model II at $z=$[6, 7, 8] have Gini coefficients of [0.93, 0.92, 0.90]. For reference, a distribution comprised of equal numbers has a Gini coefficient $\sim0$ and a distribution where all the density is held by a single member has a coefficient of $\sim0.99$. Drawing again from the language of wealth concentration, we christen this ultra-minority of galaxies that dominate reionization the ``oligarchs".

In Figure \ref{fig:fescfrac} we show the occurrence of the oligarchs grows with redshift as \sigs increases and galaxies become more compact and star-forming. Consequently, the mean \fescs also grows with redshift but note that it never exceeds an average of $\sim25\%$ even in the highest mass bin ($\log(M_{\star}/M_{\odot})=9-10$). At $z\sim4$ we predict a mean \fescs of $\sim10\%$ with $\sim10\%$ of sources displaying \fesc$>20\%$ for galaxies at $\log(M_{\star}/M_{\odot})=9-10$. This is a faithful representation of the current observational situation at $z\sim2-4$ where stacks produce average \fescs of $\sim10\%$ \citep[][]{Marchi17, Steidel18, Oesch19} while a small fraction of sources show \fescs$>20\%$, even reaching \fesc$>50\%$ \citep[e.g.,][]{Naidu17, deBarros15, Vanzella18} for galaxies in a similar mass range. In fact, in the \citet{Steidel18} sample at $z\sim3$ (which we approximately compare with our predictions at $z\sim4$) 10/124 sources ($\sim8\%$) show significant LyC leakage while the stacked mean is $\sim10\%$ -- the agreement in the fraction of sources with high \fescs is noteworthy since we fit our model against the mean \fescs and the fraction of \fesc$>20\%$ galaxies is a genuine prediction (top-right panel of Figure \ref{fig:fescfrac}).

In Model I this distribution of the ionizing budget is a direct reflection of the shape of the UVLF arising from our model since $\xi_{\mathrm{ion}}$ does not vary with \muvs (Figure \ref{fig:emissivity_zevol}) and all galaxies have the same \fesc. Steeper \auvs keeping all else same, will lead to a lower average \fesc, larger luminosity densities at early times, a less oligarchic distribution, and possibly tension with current constraints favoring late and rapid reionization (see \S\ref{sec:democratic}). However, in Model II, $M_{\mathrm{UV}}>-17$ galaxies are limited to very low \fescs and they constitute a negligible portion of the reionization budget. Truncating as high as $M_{\mathrm{UV}}=-17$ has no effect on the model parameters shown in Figure \ref{fig:model2constraint}, i.e., the ionizing emissivity between $M_{\mathrm{UV}}=-13.5$ to $-17$ is severely down-weighted by assigning a low \fesc. Even with steeper \auv, we expect the oligarch scenario to hold since Model II has the flexibility to ensure late reionization as required by the constraints by setting \fescs$\sim0$ for the numerous faint galaxies with low \sig. We discuss \auvs further in \S\ref{sec:democratic}.

\begin{figure*}
\centering
\includegraphics[width=1\linewidth]{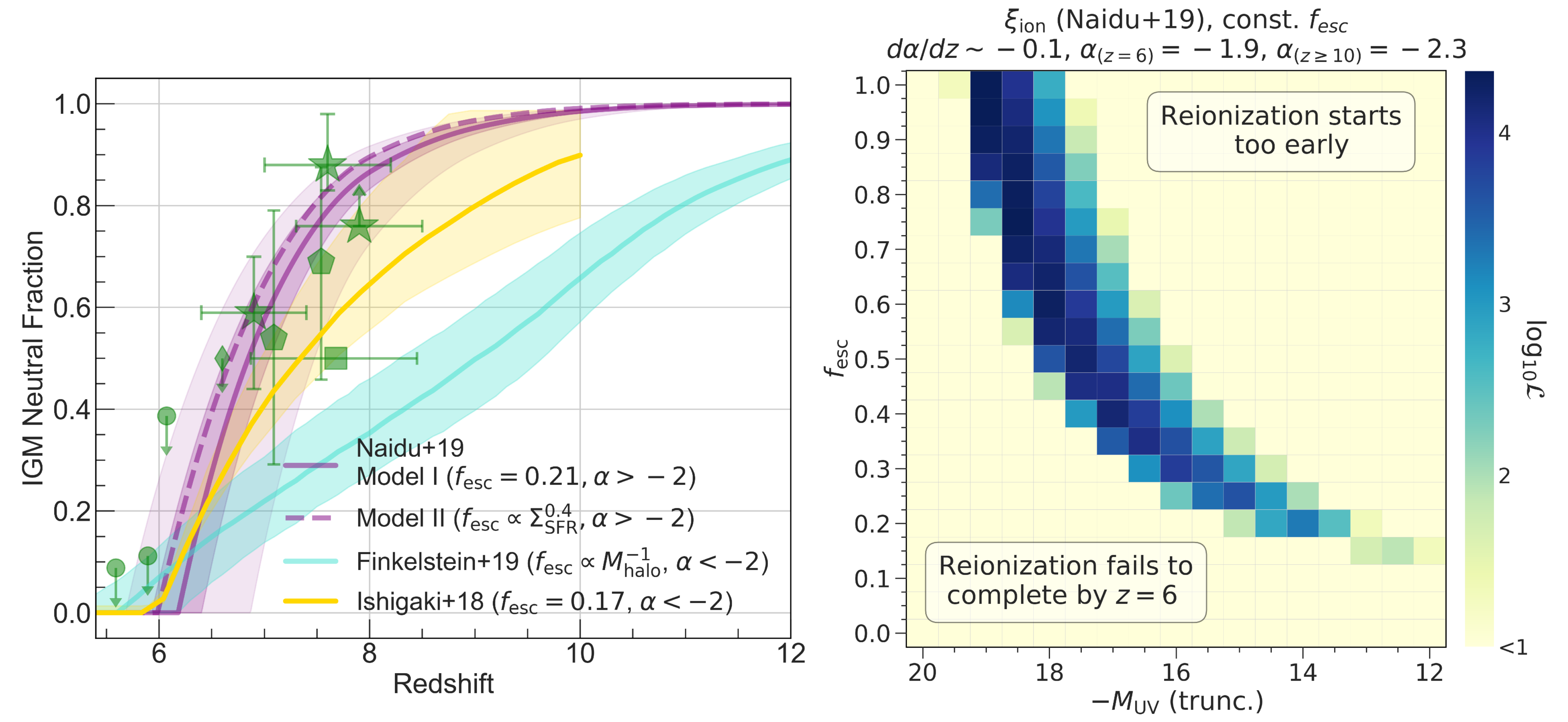}
\caption{Comparison with reionization by faint galaxies. \textbf{Left:} In turquoise we plot $\bar{x}_{\mathrm{HI}}(z)$ from \citet{Finkelstein19} who explore reionization dominated by \muv$>-15$ galaxies with steep faint-end slopes ($\alpha<-2$) and the highest \fescs occurring in the least massive galaxies by integrating down to $M_{\mathrm{UV}}=-10$. In gold we plot $\bar{x}_{\mathrm{HI}}(z)$ from \citet[][]{Ishigaki18} who assume a constant \fescs and $\alpha_{\mathrm{UV}}<-2$ to find \fesc$=17\%$ and $M_{\mathrm{UV}}\mathrm{(trunc)}=-11$ in order to complete reionization by $z=6$. Both these models ionize a large volume of the universe at early times, in tension with Ly$\alpha$ damping wing constraints (green stars and pentagons). On the other hand, the shallower faint-end slopes (\auv$>-2$) and \fescs distributions highly skewed toward bright galaxies in our models ensure rapid, late reionization (purple curves). \textbf{Right:} Assuming Schechter parameters from \citet[][]{Finkelstein19}, a constant \fescs across all galaxies, and $\xi_{\mathrm{ion}}$ from this work, we show the likelihood of various combinations of \fescs and \muv-truncation arising from the constraints in \S\ref{sec:data}. When the ionizing emissivity is dominated by faint galaxies (\muv$>-15$), even with very low \fesc, early reionization occurs, and such scenarios are disfavored compared to those starring brighter galaxies.}
\label{fig:democratic}
\end{figure*}

\subsection{``Democratic" Reionization by Faint Galaxies and the Faint-End Slope of the UVLF in a Rapidly Reionizing Universe}
\label{sec:democratic}

Reionization by oligarchs stands in sharp contrast to ``democratic" reionization that is dominated by copious faint sources that lie at \muv$>-18$ and might potentially have high escape fractions \citep[e.g.,][]{Oesch09, Bouwens11c, Wise14, Atek15, Anderson17, Livermore17, Finkelstein19}. Faint galaxies emerged as the candidate-leaders of reionization because the steep slopes (\auvs$\leq-2$ at $z>6$) of the UVLF measured after the installation of $HST$/WFC3 implied they dominated the luminosity density \citep[e.g.,][]{Bouwens12a}. The $\tau$ measurements from WMAP-9 ($0.089\pm0.014$, $z_{50}=10.5\pm1.1$) and \citet{Planck16} ($0.066\pm0.013$, $z_{50}=8.8\pm1.3$) required significant reionization at $z>8$ and hence large contributions towards the ionizing emissivity from faint galaxies \citep[e.g.,][]{Robertson13, Robertson15, Bouwens15c}. Concurrently, the very low \fescs reported for bright star-forming galaxies out to $z\sim4$ (see \S\ref{sec:introduction}) and the sharply dropping AGN luminosity function \citep{Kulkarni19} further shifted the spotlight onto faint star-forming galaxies. 

However, the recent constraints on neutral fractions detailed in \S\ref{sec:data} and the latest \textit{Planck} $\tau$ favor late, rapid reionization between $z=6-8$ (we calculate $z_{50}=6.83^{+0.24}_{-0.20}$ for Model I) i.e., high emissivity from faint galaxies at $z>8$ is no longer required. This, and the high average \fescs measured even for more massive, \muv$<-18$ galaxies allow for reionization by the oligarchs. At $z>8$, \ndots must be low enough for the universe to remain significantly neutral ($\gtrsim90\%$), and between $z=8-6$ it must rise sharply to complete reionization. Since $\xi_{\mathrm{ion}}$ evolves modestly with redshift and across \muvs (see Figure \ref{fig:emissivity_zevol}), \ndots effectively depends on $\rho_{\mathrm{SFR}}$ (\auv, \muvs truncation) and \fesc. 

Latest studies report $\alpha_{\mathrm{UV}}\lesssim-2$ at $z\geq6$, albeit with significant uncertainties, that grows steeper with redshift at a rate $d\alpha/dz\sim-0.1$ \citep[e.g.,][]{Finkelstein15,Livermore17,Bouwens17,Atek18,Ishigaki18, Oesch18}. We compare our reionization histories with models that assume these steep slopes and model $\rho_{\mathrm{SFR}}$ based on Schechter parameters extrapolated from $z<10$ fits in Figure \ref{fig:democratic}. Assuming \auv$<-2$ and setting \fescs preferentially higher in the faintest galaxies requires integration down to \muv$=-10$ and reionizes large volumes of the $z>8$ universe, reaching $\bar{x}_{\mathrm{HI}}\sim40\%$ at $z=8$ \citep[][]{Finkelstein19}, in tension with the damping wing measurements ($\bar{x}_{\mathrm{HI}}\sim90\%$). Using a constant \fescs across all galaxies with \auv$\leq-2$ like in \citet{Ishigaki18} and \citet[][]{Robertson15} requires integrating to $-$\muv$=11-13$ and still makes for too low of a neutral fraction ($\sim60-70\%$) at $z=8$. Simply lowering the constant \fescs in these models would delay reionization but then it would not conclude by $z\sim6$ -- raising the \fescs while lowering the \muvs(trunc.) and/or shallower \auvs are needed. We illustrate this in the right panel of Figure \ref{fig:democratic}, where we assume Schechter parameters from \citet[][]{Finkelstein19}, $\xi_{\mathrm{ion}}$ from this work, and a constant \fescs to evaluate how likely various combinations of \fescs and \muvs truncation are (as per constraints from \S\ref{sec:data}). A truncation \muvs of $\leq-15$, implying a limited role for fainter galaxies is favored by the constraints. The general feature of \ndots dominated by faint galaxies in the models discussed above is that \ndots is already high at $z=10$, resulting in smooth and early reionization (Figures \ref{fig:csfrd} and \ref{fig:democratic}).

On the other hand, the required late and rapid reionization is naturally produced by shallower ($\alpha_{\mathrm{UV}}\geq-2$) faint-end slopes (Model I, Model II), distributions of \fescs highly skewed toward brighter galaxies (Model II), and/or a sharp drop in the $z>8$ $\rho_{\mathrm{SFR}}$ in models linking star-formation to dark matter accretion \citep[e.g. Model I, Model II,][]{Mason15,Mashian16}. Truncating at $M_{\mathrm{UV}}=-16\ (-17)$ in Model I (Model II) produces no change in the reionization histories and model parameters, i.e., the ionizing emissivity requires no contributions from $M_{\mathrm{UV}}>-16\ (-17)$ galaxies (Figures \ref{fig:AOE} and \ref{fig:muvtrunc}). $M_{\mathrm{UV}}>-16$ galaxies are rare in the early universe and their appearance causes \ndots to rise steeply by more than a dex between $z=6-10$ (Figure \ref{fig:csfrd}). Thus, while the faint-end slope of the UVLF may indeed be extremely steep, galaxies at $M_{\mathrm{UV}}>-16$ must play only a minimal role in order to achieve rapid and late reionization. Model II explains this as very low \fescs occurring in these abundant albeit low \sigs galaxies.

\subsection{Observing the Oligarchs in Action: Promising Hints and Future Prospects}
The luminous Ly$\alpha$ emitter, COLA1 at $z=6.6$ (\fesc$\sim30\%$, \muv$=-21.5$, $\Sigma_{\rm SFR}=100\ \mathrm{M}_{\odot}/\mathrm{yr}/\mathrm{kpc}^2$) is a poster-child oligarch \citep{Hu16,Matthee18}. It displays double-peaked Ly$\alpha$ with a low-peak separation reminiscent of local LyC leakers \citep{Verhamme17}. More statistically, \citet{Songaila18, Hu16, Matthee18} find luminous Ly$\alpha$ emitters at $z\sim7$ have line profiles that are broader (while not being Active Galactic Nuclei (AGN)) and more complex than their lower luminosity counterparts with two sources in a sample of seven showing blue wings despite a highly neutral IGM \citep[e.g.,][]{Mason18}. We speculate these galaxies are oligarchs with high escape fractions that are able to carve out their own ionized bubbles perhaps allowing for their whole line profiles (including blue wings and peaks) to escape unattenuated by neutral gas. The lower luminosity, low \fescs sources have narrow, less complex Ly$\alpha$ profiles that are perhaps truncated by the neutral gas surrounding them. High-resolution ($R>4500$) Ly$\alpha$ surveys with well-defined selection and completeness functions at $z\sim0-6$ will help test if these complex Ly$\alpha$ profiles that have been linked to ionized channels and thus LyC \fescs \citep{Vanzella19,Rivera-Thorsen19, Herenz17} grow more common with redshift and with galaxy properties like \sig. Since we do not expect \fescs to evolve appreciably between $z\sim6-8$ as \sigs flattens (Figure \ref{fig:sfrsd_zevol}), such a survey can be limited to $z<6$ where the IGM transmission is higher and Ly$\alpha$ is easily observable.

Another intriguing observation is that of an overdensity of 17 \textit{HST} dropouts at $z\sim7$. In an extremely long integration (22.5 hrs on VLT/VIMOS), \citet{Castellano18} find Ly$\alpha$ emission arising only from three UV-bright galaxies among the dropouts while all their faint galaxies are undetected in Ly$\alpha$ despite Ly$\alpha$ EWs generally anti-correlating with brightness. We speculate the bright oligarchs with high \fescs have reionized their immediate surroundings rendering them transparent to Ly$\alpha$ while the fainter sources lie just outside these ionized bubbles. With \textit{JWST}'s planned censuses at high-$z$ more such ionized overdensities at $z>6$ will come into view and deep follow-up spectroscopy that reveals features of LyC \fescs (e.g., multi-peaked Ly$\alpha$) will help test if they are indeed powered by oligarchs.

Our proposed scenario also has strong implications for the topology of reionization, with the distribution of ionized bubble sizes and the patchiness resulting from our model likely lying somewhere intermediate between AGN-driven reionization \citep[e.g.,][]{Kulkarni17} and reionization by widely distributed, numerous faint sources \citep[e.g.,][]{Geil16}. Upcoming 21cm surveys will provide a strong test of this prediction \citep[e.g.,][]{Hutter19b,Hutter19a,Seiler19}. Our empirical model also tracks the spatial distribution of galaxies and this information can be coupled with models for \fescs to produce more quantitative predictions. We defer this to future work. 

\subsection{Related Work: the \fesc-\sigs Connection}
\citet{Heckman01} explicitly link \fescs to a critical value of \sigs$\sim \mathrm{0.1 M_{\odot}/yr/kpc^{2}}$ above which they observe the occurrence of strong winds becomes common in star-burst galaxies. They hypothesize that these winds are responsible for LyC leakage. \citet{Sharma16} adopt this idea in the \texttt{EAGLE} simulations \citep{Schaye15,Crain15} setting \fesc$=0.2$ for star-forming regions above the critical \sigs and averaging these in each galaxy to find an emissivity consistent with \citet{Bouwens15c}. The \fesc$=0.2$ hard upper-limit was motivated by the lack of LyC detections, prior to the recent LyC renaissance detailed in \S\ref{sec:introduction}. 

We do not assume a threshold or an upper bound emboldened by recent discoveries (see Figure \ref{fig:sfrsd_zevol}) and empirically constrain an \fesc-\sigs dependence. We use a prior that allows for no relation ($b=0$) or an anti-correlation ($b<0$) and \textit{fit} against the latest reionization constraints. Note that the \citet{Sharma16} prescription, even though it invokes \sigs, ends up effectively similar to our Model I that fits \fesc=$0.21^{+0.06}_{-0.04}$, since essentially all galaxies at $z\sim6-8$ at $M_{\mathrm{UV}}<-13.5$ have \sig$>\mathrm{0.1 M_{\odot} yr^{-1}kpc^{-2}}$ (see bottom left panel in Figure \ref{fig:model1vmodel2}) and have sizes on the order of $1$ kpc (their beam size). Furthermore, \citet{Sharma16} find \texttt{EAGLE} galaxies at $M_{\mathrm{UV}}<-18$ produce $\sim50\%$ of the reionization budget. Our budget is far more oligarchic ($>80$\%), and our reionization far more rapid ($z\sim6-8$; see \S\ref{sec:rapid}).

\section{Caveats and Open Questions}
\label{sec:caveats}

\subsection{On Dust and \fescs at $z>6$}
\label{sec:dust}
In our Model I, \fescs implicitly folds in the role of dust and all other processes that may curtail LyC leakage. This is not the case in Model II in which \fescs is solely a function of $\Sigma_{\mathrm{SFR}}$. Further, we adapt the \citet{Steidel18} measurement to constrain Model II assuming that the relationship between \sigs and \fescs at $z\sim3$ carries over to higher-$z$, unmodulated by dust. We justify these assumptions here.

At $z\sim3$ already, deep stacks of typical LBGs show that it is not dust, but photoelectric absorption in the ISM that dominates the attenuation of LyC photons to the extent that
$$f_{\mathrm{esc}} \approx 1 - f_{\mathrm{cov}}$$ 
where $f_{\mathrm{cov}}$ is the \ion{H}{1} covering fraction \citep{Reddy16a,Reddy16b,Steidel18}. Moving into the EoR, this approximation is likely even better since the dust attenuation at $z>6$ appears to be lower \citep[][]{Bouwens16, Fudamoto17} though significant uncertainties persist \citep[e.g.,][]{Casey18}. Our attenuation prescription bears this out -- for instance, on average, an $M_{\mathrm{UV}}=$[$-22$,$-20$,$-18$] galaxy at $z=4$ has $A_{\mathrm{UV}}$=[1.51, 1.05, 0.6] and a $z=8$ galaxy has $A_{\mathrm{UV}}$=[1.40, 0.6, 0.0]. The key physical picture motivating Model II, that of spatially concentrated star-formation, winds, and feedback carving out ionized gas channels is intimately linked with $f_{\mathrm{cov}}$ and thus needs no extra dust parameter, since $f_{\mathrm{cov}}$ essentially determines \fesc. However, we note that not explicitly modeling dust in Model II prevents a simple extrapolation of \fescs using our fit power-law to $z\sim0$ where attenuation of LyC by dust is highly significant though the qualitative picture and general trend stands.

\subsection{Effect of $M_{UV}$ Truncation}
\label{sec:muv_trunc}
\begin{figure}
\centering
\includegraphics[width=1\linewidth]{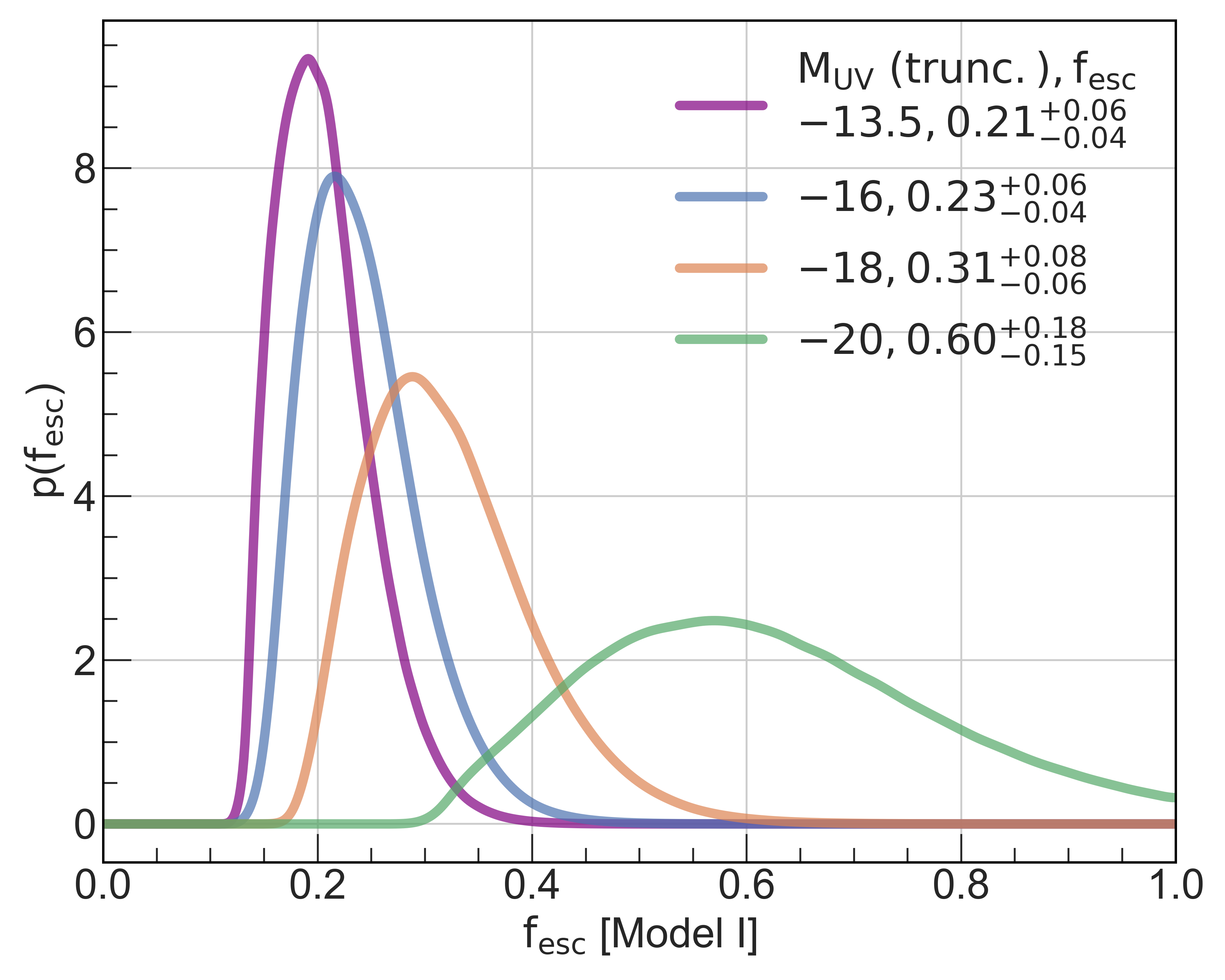}
\caption{The effect of \muvs truncation on our \fescs posterior from Model I. \fescs converges as we go to lower \muv. We do not expect any significant change from extensions to even fainter magnitudes as the fractional increase in the cumulative \ndots becomes negligible at \muv$>-16$. This is expected since our model produces \auv$>-2$, and thus convergent cumulative luminosity densities, during the EoR.}
\label{fig:muvtrunc}
\end{figure}

We have limited all our calculations to galaxies with $\mathrm{SFR}>0.02\ \mathrm{M_{\odot}}\mathrm{yr^{-1}}$ which corresponds approximately to \muv(observed)$<-13.5$. This limitation arises from the resolution of the dark matter simulations our model is built on as well as the significant uncertainties around the UVLF fainter than this magnitude \citep{Livermore17, Bouwens17, Atek18}. What effect does this truncation have on our results?

In Model I extending to fainter magnitudes adds to the ionizing emissivity and should lower the average \fescs we report. However, since our model has \auv$>-2$ during reionization we expect this lowering to become negligible at \muv$>-13.5$, since the differential change to the bulk \ndots asymptotes to zero. We explore this by shifting the limiting magnitude brighter (see Figure \ref{fig:muvtrunc}). We find our \fescs solution is essentially converged at \muv$<-16$ since moving down to \muv$<-13.5$ produces no appreciable change. Moving further down to \muv$<-11$ should make an even smaller difference especially if the UVLF turns over around these magnitudes due to inefficient star-formation and photo-evaporation in low-mass halos \citep[e.g.,][]{Gnedin16}. 

In Model II the majority of the low-luminosity galaxies have extremely low \fescs (see bottom panels of Figure \ref{fig:fescfrac}) to go along with their lower LyC output so the exclusion of \muv$>-13.5$ galaxies or the faint-end slope have negligible impact on model parameters. We have verified that even truncating as high as \muv$<-17$ produces very similar reionization histories to those reported here.

\subsection{Model Dependent Constraints}

The Ly$\alpha$ damping measurements for $z>7$ quasars and galaxies prove to be the most constraining for our Model I. However, these are model dependent constraints in that their reported $\bar{x}_{\mathrm{HI}}$ is a product of several assumptions, e.g., about how Ly$\alpha$ is processed by the ISM at high-$z$ or about the intrinsic spectrum of reionization epoch quasars. These assumptions while reasonable are yet to be tested \citep[e.g., see \S2 in][]{Mason18}. In Model I the model-independent $\tau$ and dark fraction by themselves are unable to zero in on an \fescs solution, but they rule out \fesc$\lesssim15\%$ and so favor rapid reionization histories. In Model II the Ly$\alpha$ damping measurements are unable to collapse the posterior much further beyond $\tau$ and the dark fraction combined, and the \citet{Steidel18} measurement proves crucial (see Figure \ref{fig:model2constraint}). This measurement depends on several assumptions e.g., stellar population model predictions at $<912\mathrm{\AA}$, the IGM+CGM transmission functions, and details of the ``hole" and ``screen" models for \fescs developed in their work. These are all sources of systematic uncertainty on the reported absolute \fesc. We have tested that the sign of the power ($>0$) recovered for the \fesc-\sigs dependence is not sensitive to the exact scale of their reported \fescs as long as \fesc$>0$, and the dark fraction measurement that ensures the timely conclusion of reionization is used.

\subsection{Ionizing Emissivity at $z<6$ and the Role of AGN}
\label{sec:AGN}
In this work we have focused on reionization driven by galaxies and fit for parameters that show them satisfying all constraints outlined in \S\ref{sec:data} without invoking AGN. This is supported by latest determinations of AGN luminosity functions that limit their contribution to $<5\%$ at $z=6$ \citep[e.g.,][though see \citealt{Giallongo15,Boutsia18}]{Kulkarni19}. In a companion work we deploy a similar framework, but assume nothing about the underlying ionizing population and fit a non-parametric model to recover \ndot$(z)$ \citep{Mason19}. A sharp drop in \ndot$(z)$ at $z=6-8$ fully consistent with the galaxy-only models presented in this work is recovered.

A related issue is whether our models overrun the constraints on ionizing emissivity at lower-$z$ plotted in the top-left panel of our Figure \ref{fig:model1vmodel2} when combined with the AGN emissivity at lower redshifts \citep{Becker13, Kuhlen12}. Note that in Model I we make no claims about \fescs at $z<6$ and fit a $\sim20\%$ \fescs during the short window when reionization transpires. In Model II we have an evolving \fescs that falls from $\sim20\%$ at $z>6$ to $\sim10\%$ ($\sim4\%$) at $z=3$ ($z=2$). This causes \ndots to flatten and turn-over at lower-$z$ so that AGN can dominate the emissivity -- we begin to see this in the top-left panel of Figure \ref{fig:model1vmodel2} (dotted purple curve).

\subsection{Cosmic Variance and Completeness}
\label{sec:cosmicvar}

Due to the finite volume of the N-body simulation on which our empirical model is built ($10^{6}$ Mpc$^{3}$), we miss some of the most massive halos (see Figure 20 and Appendix B in \citealt{Tacchella18}). At $z>6$ these halos also tend to be the most star-forming and UV bright. Comparing to an analytical halo mass function \citep{Sheth01} and applying a completeness correction produces a $\lesssim0.3$ dex difference at the brightest end of the UVLF at $z=10$. The correction is smaller at lower redshifts, where the bulk of reionization occurs and hence the magnitude of the effect is likely small. For Model I, the mean \fescs estimated would slightly shrink due to the missing luminosity density. For Model II, extrapolating from the trends shown in Figure \ref{fig:fescfrac} for the proportion of oligarchs as a function of galaxy mass at $z>6$, including these massive halos would make the reionization budget even more oligarchic. An update to the empirical model using a larger box that also self-consistently includes AGN is currently under preparation. The larger box will also allow us to address cosmic variance by resampling smaller volumes and computing the resulting scatter introduced in \fesc.

\section{Summary}
\label{sec:summary}
Using an empirical model that accurately predicts observations (e.g., the sharp drop in $\rho_{\rm SFR}$ at $z>8$, UVLFs at $z>4$, $\xi_{\mathrm{ion}}$ at $z\sim4-5$, the $z$-evolution of \sig) and leveraging recent measurements of the timeline of reionization (e.g., $\bar{x}_{\mathrm{HI}}$ at $z\gtrsim7$, the \textit{Planck} $\tau$) we constrain \fesc, the most uncertain parameter in reionization calculations. Deploying two models -- one assuming a constant \fescs across all galaxies during reionization (Model I), another linking \fescs to \sigs (Model II) -- we find the following:

\begin{itemize}
    \item In both our models, \muv$<$$-$13.5 star-forming galaxies need an average \fesc$\sim$20$\%$ at $z>6$ to conclude reionization, a factor of only 2 higher than the recently measured \fesc$\sim10\%$ at $z\sim2.5-4$. [Figures \ref{fig:fourpanel}, \ref{fig:model1vmodel2}]
    \item Our Model II explains this evolution in \fescs by appealing to \sigs that decreases by $\sim$2.5 dex between $z=8$ (\fesc$\sim0.2$) and $z=0$ (\fesc$\sim0$) by fitting \fesc$\propto($\sig$)^{0.4}$. This \fesc-\sigs connection is inspired by the newly emerging sample of LyC leakers that show higher \sigs than the average galaxy population at their redshifts. Latest hydrodynamical simulations qualitatively support the idea of spatially concentrated star-formation blowing channels in the ISM through which LyC produced by long-lived, rotating, binary stars escapes. [Figures \ref{fig:sfrsd_zevol}, \ref{fig:model2constraint}]
    \item The universe goes from $90\%$ neutral to $10\%$ neutral in a short span of $\sim300$ Myrs between $z\sim6-8$, and favored by Ly$\alpha$ damping measurements requiring a $\gtrsim50\%$ neutral universe at $z\sim7$. This conclusion stands even considering only model-independent constraints ($\tau$, dark fraction) that rule out \fesc$\lesssim15\%$. [Figure \ref{fig:fourpanel}, Table \ref{table:rapid}]
    \item The bulk of the reionization budget ($\sim$50$\%$ in Model I, $\sim$80$\%$ in Model II) is concentrated among a small number ($<$5$\%$) of galaxies (the ``oligarchs"). This is due to the faint-end slopes of the UVLF (\auv$>-2$) in our model and the distribution of \fescs skewed toward high \sig, massive galaxies. The oligarchs are compact ($R_{\rm gal}\sim0.5$ kpc), have higher \sigs than average ($\sim$10$-$20 $\mathrm{M_{\odot} yr^{-1} kpc^{-2}}$), are relatively massive ($\log(M_{\star}/M_{\odot})$$>$8) and are UV bright (\muv$<$$-$18). The fraction of these oligarchs grows with redshift, while keeping the average \fescs to $\sim$20$\%$. Extrapolating to $z\sim3-4$ we match the current situation where a small fraction of galaxies ($\lesssim$10$\%$) display \fesc$>0.2$, some even exceeding $50\%$, while the average \fescs stays at $\sim$10$\%$. [Figures \ref{fig:fescfrac}, \ref{fig:AOE}, Table \ref{table:AOE}]
    \item Faint galaxies are disfavored to drive reionization. When faint galaxies with steep $\alpha_{\mathrm{UV}}<-2$ dominate the emissivity, they ionize large volumes of the universe at $z=7-8$, in tension with Ly$\alpha$ damping constraints that require a $60-90\%$ neutral universe at these redshifts. Shallower faint-end slopes ($\alpha_{\mathrm{UV}}>-2$) and/or \fescs distributions skewed toward massive galaxies like in our models ensure high neutral fractions at late times while also completing reionization by $z=6$. Concurrently, the motivation for excluding galaxies at $M_{\mathrm{UV}}<-18$ with high \fescs as the protagonists of reionization has grown weaker as the observational picture has shifted to these galaxies being able to produce \fescs$\sim10\%$ at $z=2.5-4$. [Figure \ref{fig:sfrsd_zevol}, \ref{fig:democratic}]
\end{itemize}

Our predictions are eminently testable since the oligarchs are bright, currently observable galaxies. Deep Ly$\alpha$ surveys at high-resolution ($R>4500$) spanning $z\sim0-6$ should show a growing incidence of galaxies with multi-peaked Ly$\alpha$ at higher-$z$. These peaks represent ionized channels for LyC escape, as seen in the $z\lesssim4$ LyC leakers. Upcoming 21 cm experiments should infer a bubble size distribution with a high Gini coefficient as the first ionization fronts form predominantly around the oligarchs.

\acknowledgments{We are grateful to the referee, Brant Robertson, for a thorough report that greatly improved the clarity of this work. It is a pleasure to acknowledge illuminating discussions with Steve Finkelstein, Jorryt Matthee, Daniel Schaerer, David Sobral, Irene Shivaei, Ben Johnson, and Chuck Steidel that enriched this work. We thank Daniel Eisenstein, Avi Loeb, and Lars Hernquist for participating in a robust discussion of these results during RPN's research exam. We thank Steve Finkelstein for sharing the posteriors from \citet{Finkelstein19} featured in Figure \ref{fig:democratic}. We are grateful to Eros Vanzella for sharing their data on \textit{Ion3} and for useful comments on \citet{Naidu18} that sparked parts of this work. RPN thanks Michelle L. Peters for a hilarious proofreading session that highlighted the ludicrous verbiage we deploy in the astrophysics community. RPN thanks the organizers and attendees of IAU Symposium 352 ``Uncovering early galaxy evolution in the ALMA and JWST era" held in Viana do Castelo for providing much-needed affirmation that he is not alone in the often testing endeavour that is working on \fesc.

RPN gratefully acknowledges an Ashford Fellowship and Peirce Fellowship granted by Harvard University. ST is supported by the Smithsonian Astrophysical Observatory through the CfA Fellowship. CAM acknowledges support by NASA Headquarters through the NASA Hubble Fellowship grant HST-HF2-51413.001-A awarded by the Space Telescope Science Institute, which is operated by the Association of Universities for Research in Astronomy, Inc., for NASA, under contract NAS5-26555. SB acknowledges Harvard University's ITC Fellowship. PAO acknowledges support from the Swiss National Science Foundation through the SNSF Professorship grant 157567 ``Galaxy Build-up at Cosmic Dawn". The Cosmic Dawn Center is funded by the Danish National Research Foundation. CC acknowledges funding from the Packard foundation.}

\software{
    \package{IPython} \citep{ipython},
    \package{matplotlib} \citep{matplotlib},
    \package{numpy} \citep{numpy},
    \package{scipy} \citep{scipy},
    \package{jupyter} \citep{jupyter},
    \package{seaborn} \citep{seaborn},
    \package{FSPS} \citep{fsps1, fsps2},
    \package{dynesty} \citep{Speagle19},
    \package{python-FSPS} \citep{python-FSPS}
    }

\bibliography{MasterBiblio}
\bibliographystyle{apj}

\appendix
\section{UV Luminosity Functions}
\label{appendix:uvlf}
In addition to Figure \ref{fig:uvlf}, where we compare against literature data at $z=6$ (the redshift at which the faint-end of the UVLF has been measured), we provide UVLFs predicted by our empirical model for $z=7-12$. We also quantify agreement between our model and observational data in Table \ref{table:chisq}. In general, our predicted UVLFs show good agreement with the data, both at the bright and faint ends. The only tension is at $z=6$ with \citet{Atek18} at $M_{\mathrm{UV}}<-16$ where they estimate several more galaxies than predicted by our model -- we find better agreement with \citet{Bouwens15aLF} and \citet{Livermore17} in the same luminosity regime, while also noting that the excess of brighter galaxies would further support this paper's point of view about the reionization budget.

\begin{figure}[h]
\centering
\includegraphics[width=0.5\linewidth]{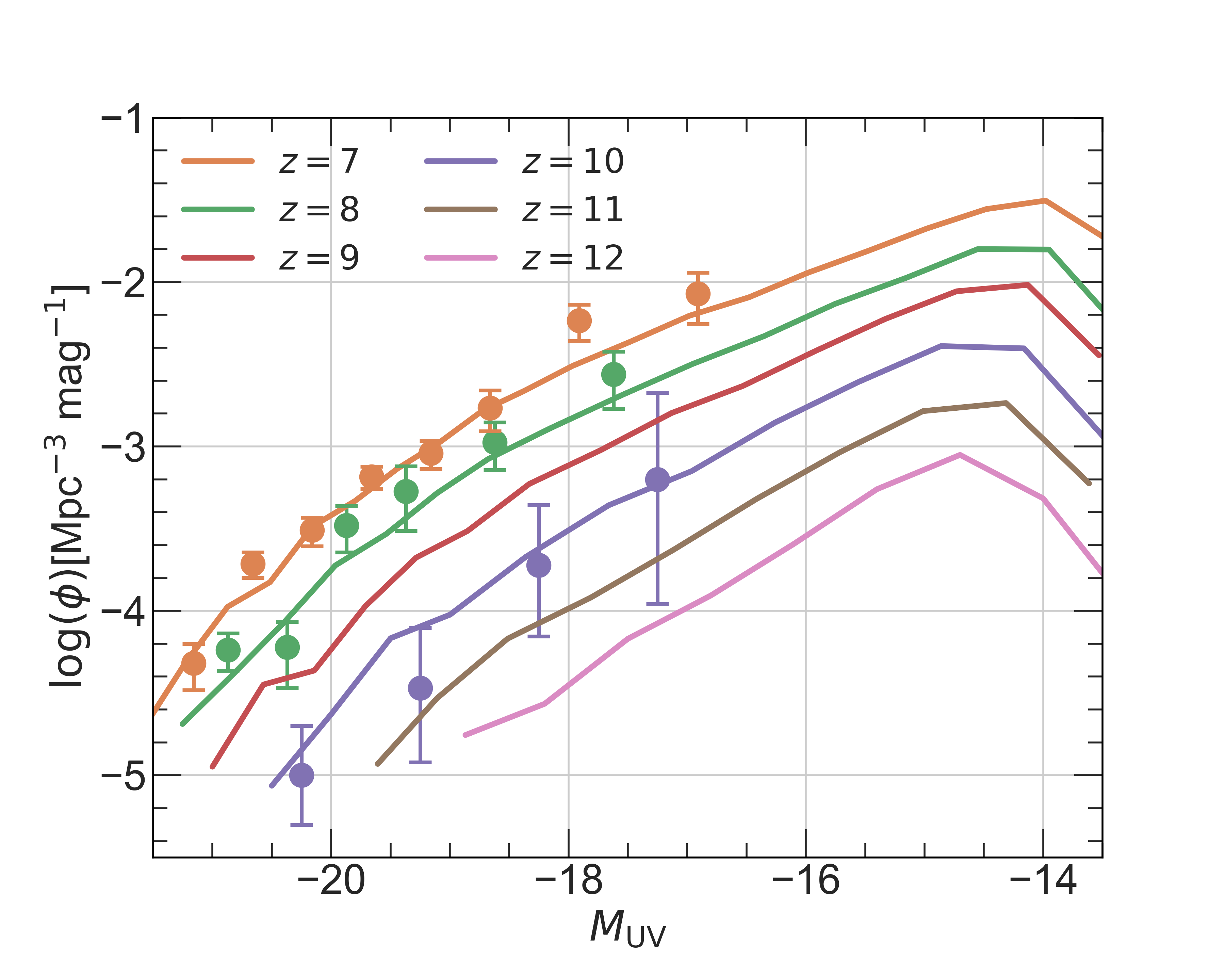}
\caption{UVLFs from the \citet[][]{Tacchella18} model compared with data from \citet[][]{Bouwens15aLF} ($z=7-9$) and \citet[][]{Oesch18} ($z=10$).}
\label{fig:uvlfz712}
\end{figure}

\begin{table*}[h]
\centering
\caption{Comparison of Model UVLFs against literature data.}
\label{table:chisq}
\begin{tabular}{lccc}
\hline
\multicolumn{1}{l}{Redshift} & \multicolumn{1}{c}{$\chi^{2}$ ($M_{\mathrm{UV}}<-16$)} &
\multicolumn{1}{c}{$\chi^{2}$ ($M_{\mathrm{UV}}>-16$)}&
\multicolumn{1}{c}{Source}\\
\hline
\noalign{\smallskip}
$\mathrm{6}$ & $0.7$ &
$0.5$ &
\citet{Bouwens17}\\
$\mathrm{}$ & $9.9$ &
$0.1$ &
\citet{Atek18}\\
$\mathrm{}$ & $1.5$ &
$1.7$ &
\citet{Livermore17}\\
$\mathrm{7}$ & $1.4$ &
$-$&
\citet{Bouwens15c}\\
$\mathrm{8}$ & $1.3$ &
$-$ &
\citet{Bouwens15c}\\
$\mathrm{10}$ & $0.9$ &
$-$ &
\citet{Oesch18}\\
\noalign{\smallskip}
\end{tabular}
\begin{tablenotes}
\centering
\item[](a) $\chi^{2}$ is reported as $\frac{1}{n-1}\Sigma_{i=1}^{n}\frac{(x_{i}-\mu_{i})^{2}}{\sigma_{i}^{2}}$, where $x$ and $\mu$ represent the data and model values respectively, and $\sigma$ is the error on $x$. When reported errors are non-Gaussian, we set $\sigma$ to half the difference between the 84th and 16th percentile.
\end{tablenotes}
\end{table*}

\end{document}